# Quickest Detection of Dynamic Events in Networks

Shaofeng Zou, Venugopal V. Veeravalli, Jian Li, and Don Towsley

**Abstract**

The problem of quickest detection of dynamic events in networks is studied. At some unknown time, an event occurs, and a number of nodes in the network are affected by the event, in that they undergo a change in the statistics of their observations. It is assumed that the event is dynamic, in that it can propagate along the edges in the network, and affect more and more nodes with time. The event propagation dynamics is assumed to be unknown. The goal is to design a sequential algorithm that can detect a "significant" event, i.e., when the event has affected no fewer than $\eta$ nodes, as quickly as possible, while controlling the false alarm rate. Fully connected networks are studied first, and the results are then extended to arbitrarily connected networks. The designed algorithms are shown to be adaptive to the unknown propagation dynamics, and their first-order asymptotic optimality is demonstrated as the false alarm rate goes to zero. The algorithms can be implemented with linear computational complexity in the network size at each time step, which is critical for online implementation. Numerical simulations are provided to validate the theoretical results.

## 1 Introduction

In the problem of quickest change detection (QCD), a stochastic system is observed sequentially. At some unknown time, a change occurs that changes the data generating process. Observations are taken sequentially with time, and the objective is to detect the change as quickly as possible subject to false alarm constraints (see [2–5] for an overview). The QCD framework models a wide range of applications, e.g., fraud detection, intrusion detection, environmental monitoring, line outage detection in power systems, quality control in online manufacturing systems and spectrum monitoring in wireless communications. However, in many applications, e.g., epidemic detection [6, 7], opinion mining in social networks [8], anomalous event detection in sensor networks (e.g., internet of battlefield things) [9], detection of malicious code spreading in computer networks [10],

---

The material in this paper was presented in part at the IEEE International Conference on Acoustics, Speech, and Signal Processing, Calgary, Alberta, Canada, in 2018 [1].

Research reported in this paper was sponsored in part by the Army Research Laboratory under Cooperative Agreement W911NF-17-2-0196. The views and conclusions contained in this document are those of the authors and should not be interpreted as representing the official policies, either expressed or implied, of the Army Research Laboratory or the U.S. Government. The U.S. Government is authorized to reproduce and distribute reprints for Government purposes notwithstanding any copyright notation here on. The work of S. Zou and V. V. Veeravalli was also supported in part by the National Science Foundation (NSF) under grant CCF 16-18658, and by the Air Force Office of Scientific Research (AFOSR) under grant FA9550-16-1-0077.

Shaofeng Zou is with the Coordinated Science Laboratory, University of Illinois at Urbana-Champaign, Urbana, IL 61801 USA (email: szou3@illinois.edu).

Venugopal V. Veeravalli is with the Coordinated Science Laboratory, ECE department, Department of Statistics, University of Illinois at Urbana-Champaign, Urbana, IL 61801 USA (email: vvv@illinois.edu).

Jian Li is with the College of Information and Computer Sciences, University of Massachusetts Amherst, Amherst, MA, 01003, USA (e-mail: jianli@cs.umass.edu).

Don Towsley is with the College of Information and Computer Sciences, University of Massachusetts Amherst, Amherst, MA, 01003, USA (e-mail: towsley@cs.umass.edu).

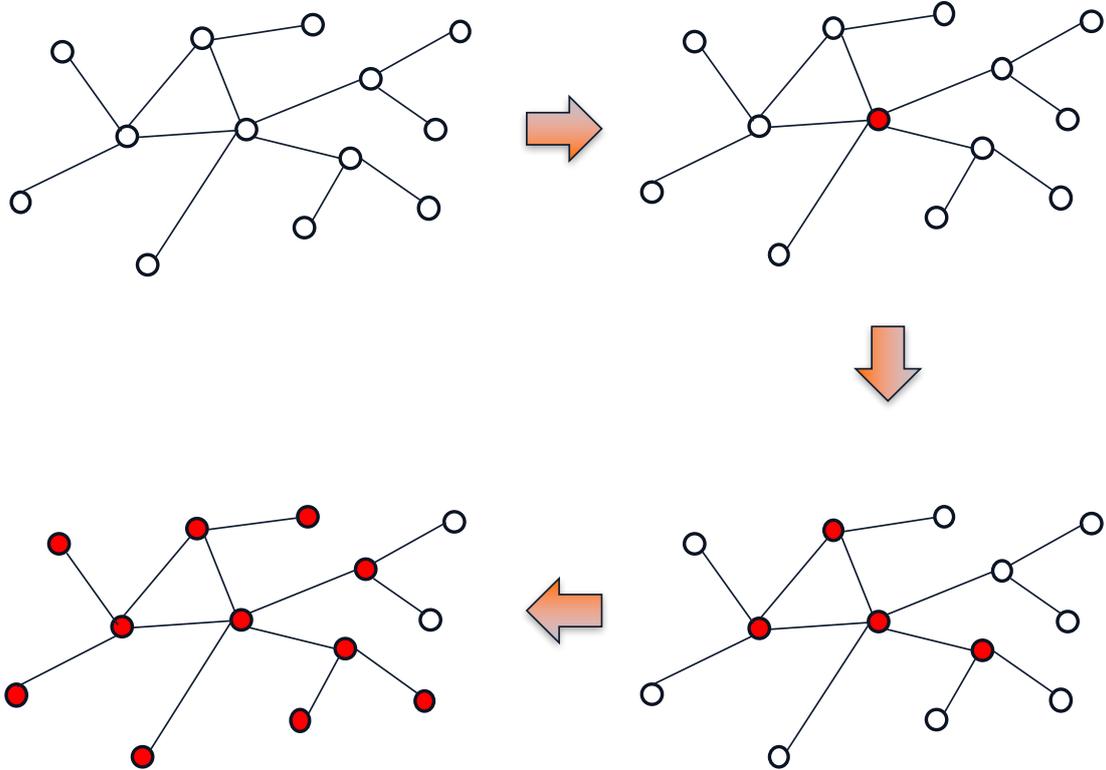

Figure 1: A dynamic event propagates in a network with time.

the data are usually collected from networks with certain underlying topologies. Following the occurrence of an event, it propagates dynamically across the network, affects more and more nodes, and changes their data generating behaviors with time (see Fig. 1). The propagation dynamics is usually unknown in practice, and depends on the underlying network topology.

Motivated by these applications, we study the problem of quickest detection of dynamic events in networks. Suppose a network is monitored in real time by a set of $L$ nodes that communicate with a fusion center. At some unknown time, an event occurs in the network that causes eventual changes in the observations of a connected subset of nodes. The event occurs at a connected subset of nodes and then dynamically propagates along the edges in the network, and the affected nodes form a connected sub-graph, the size of which grows with time. The propagation dynamics are assumed to be unknown, i.e., the set of nodes and the order in which they are affected are unknown. We are interested in detecting a "significant" event, i.e., one that affects $\eta \geq 1$ nodes as quickly as possible, subject to false alarm constraints.

## 1.1 Related Works

The problem in this paper is closely related to the problem of QCD under the multi-channel setup, in which one or multiple unknown nodes perceive a change simultaneously [11–15], or alternatively, at different times [16–18]. The major differences from these previous works lie in that: (i) we are interested in detecting whether the event has affected at least $\eta$ nodes, i.e., the event is "significant" enough, whereas previous works focus on the special case with $\eta = 1$, i.e., whether the event has occurred or not; (ii) instead of considering the worst-case performance over all possible times that the nodes are affected [18] or taking a Bayesian approach [16, 17], we assume that the times that the nodes are affected are deterministic and unknown, and we are interested in designing algorithms that adapt to unknown propagation dynamics; and (iii) we consider *structured* networks, over which events can only propagate along network edges.

On a temporal scale, the data generating distribution of the whole network dynamically changes over time. As the event affects more nodes with time, the network goes through multiple transient phases in which the sets of affected nodes are different. This is related to the problem of QCD under transient dynamics [19–21], where after an event occurs, the pre-change distribution does not change to a persistent post-change distribution instantaneously, but only after a number of transient phases. Each transient phase is associated with a distinct data generating distribution. In [19–21], it is assumed that the number of transient phases and the data generating distributions associated with each phase are known. In this paper, event propagation dynamics are unknown. Therefore, the results in [19–21] cannot be directly applied to solve the problem here. Moreover, in [20] a Bayesian approach is employed, where it is assumed that the transient durations are geometrically distributed with statistics known to the decision maker. In contrast to [20], in this paper, we make no probabilistic assumptions on the times at which the nodes are affected.

The data generating distributions for the whole network before and after $\eta$ nodes are affected are both composite, i.e., belong to a set of distributions, as they are determined by the unknown subset of affected nodes with unknown change times. Therefore, the problem in this paper is related to the problem of QCD with composite pre-/post-change distributions [22–25]. However, our problem differs from these works in the following ways. First, the data generating distribution before $\eta$ nodes are affected is composite, whereas in [22, 23, 25], only the post-change distribution is composite. Second, our test statistics can be computed efficiently at each time step with a computational complexity linear in the network size (number of nodes), whereas a sliding window approach is usually used to control the computational cost in [22–25]. Third, in our problem, the distribution of samples before and after the nodes are affected by the event are arbitrary (not necessarily belonging to an exponential family), and the parameters of the data generating distribution for the whole network (times at which the nodes are affected by the event) are discrete, and do not necessarily belong to a compact parameter space.

The offline setting of our problem has been extensively studied in the literature [26–34]. The aim of these works is to detect whether there exists a subset of nodes in the network, which have certain geometric structures (e.g., connected subgraphs), and the observations received by this subset of nodes are generated from a distribution different from the one that generates the samples for the rest of the nodes in the network. Our problem is the online dynamic version with a growing anomalous geometric structure. Hence, a small computational complexity at each time step is important in order for making timely decisions. However, many previous works are based on the scan statistic that scans over all connected subgraphs, which is computationally inefficient for large networks, and thus cannot be directly applied to our online setting. As will be shown later, our algorithms can be updated recursively with computational complexity linear in the network size at each time step, and we do not reprocess the previous data over and over again.

## 1.2 Contributions

In this paper, we start with fully connected networks, and then extend to arbitrarily connected networks.

For fully connected networks, the event can propagate from any node to any other node. Then, the algorithm design does not need to account for the fact that the event only propagates along the edges in the network, and the network structure does not matter. This simplifies the problem to one where we are simply interested in detecting when an arbitrary subset of $\eta$ nodes has been affected by the event.

For fully connected networks, we solve the QCD problem by reformulating it as a dynamic composite hypothesis testing problem, where we distinguish between two hypotheses at each time instant. The null hypothesis corresponds to the case that less than $\eta$ nodes are affected; and the alternative hypothesis corresponds to the case that at least $\eta$ nodes are affected. The data generating distributions for the whole network before and after $\eta$ nodes are affected are both composite, as they are determined by the unknown subset of affected nodes with unknown change times. We take the generalized log-likelihood ratio between the two composite hypotheses as the detection statistic, and compare it to a positive threshold. If it is greater than the threshold, then we stop and raise an alarm; otherwise, we take another sample from each node in the network.

We show that the generalized log-likelihood ratio test is equivalent to one that compares the sum of the smallest $L - \eta + 1$ local Cumulative Sum (CuSum) statistics [35] to the same threshold. The resulting algorithm, which we refer to as Spartan-CuSum (S-CuSum), is computationally efficient with $O(L)$ complexity at each time step. This guarantees that the algorithm can be implemented efficiently in an online fashion for large networks. We further show that the S-CuSum algorithm satisfies the false alarm constraints with a properly chosen threshold for all scenarios with fewer than $\eta$ affected nodes, and adapts to unknown event propagation dynamics. We then establish the asymptotic optimality of the S-CuSum algorithm up to a first-order approximation as the false alarm rate goes to zero.

For arbitrarily connected networks, the S-CuSum is still applicable. However, it does not account for the fact that the event only propagates along the edges in the network, and thus will trigger more false alarms. A direct generalization of the generalized log-likelihood ratio test involves a complicated statistic that scans over all connected sub-graphs and propagation dynamics at each time step. This is computationally intractable for a large arbitrarily connected network, especially under the online setting. We then construct an algorithm based on a thresholding approach, the breadth-first search (BFS) algorithm and the S-CuSum algorithm, which we refer to as the Network-CuSum (N-CuSum) algorithm. We show that the computational complexity of the N-CuSum algorithm is linear in the network size at each time step, and that it is asymptotically optimal up to a first-order approximation as the false alarm rate goes to zero. Moreover, we show through our numerical results that the N-CuSum algorithm, which accounts for the network structure, has a better performance than the S-CuSum algorithm. Both the S-CuSum algorithm and the N-CuSum algorithm are better than the generalized multi-chart CuSum algorithm [36], which stops when at least $\eta$ local CuSum statistics cross their individual thresholds, and the network generalized multi-chart CuSum algorithm, which stops when the local CuSum statistics of at least $\eta$ connected nodes cross their individual thresholds simultaneously.

## 1.3 Paper Organization

This paper is organized as follows. In Section 2, we formulate the problem mathematically. In Section 3, we focus on fully connected networks, present the S-CuSum algorithm, and present a method to choose the threshold to satisfy the false alarm constraints. In Section 4, we demonstrate the asymptotic optimality of the S-CuSum algorithm. In Section 5, we study arbitrarily connected networks, present the N-CuSum algorithm and demonstrate its asymptotic optimality. In Section 6, we present numerical results. Finally, in Section 7, we discuss some potential extensions.

## 2 Problem Formulation

Consider a network monitored in real time by a set of $L$ nodes. We use an unweighted, undirected graph $G = (V, E)$ to denote the underlying structure of the network. Here, $L = |V|$. In practice, an edge connecting two nodes may be due to the fact that two nodes communicate with each other, or are geometrically close to each other.

Before an event occurs, node $i \in \{1, 2, \ldots, L\}$ receives independent and identically distributed (i.i.d.) samples from distribution $f_0$. If an event occurs, and node $i$ is affected by the event at an unknown time $\nu_i$, then it starts to receive i.i.d. samples from distribution $f_1$, i.e., $\nu_i$ is the change-point at node $i$. If $\nu_i = \infty$, node $i$ will not be affected by the event ever. More specifically, if we denote the observation received by node $i$ at time $k$ by $X_i[k]$, then

$$X_i[k] \sim \begin{cases} f_0, & \text{if } k < \nu_i, \\ f_1, & \text{if } k \geq \nu_i. \end{cases} \tag{1}$$

We assume that the event first affects a *connected* subset of nodes, which might be due to the locality of the event, and then dynamically propagates along the edges in the network (see Fig. 1 for an example). Equivalently, at every time step, the induced sub-graph on all the affected nodes is *connected*.

We consider a centralized setting in which a fusion center obtains the samples of all the nodes without delay. We are interested in sequentially detecting a "significant" event, i.e., one that affects at least $\eta \geq 1$ affected nodes. If an alarm is triggered at a time when fewer than $\eta$ nodes are affected, it is then considered as a false alarm event.

Let $\boldsymbol{\nu} = \{\nu_1, \ldots, \nu_L\}$, which is unknown in advance. Without loss of generality, we assume that $\nu_1 \leq \nu_2 \leq \cdots \leq \nu_L$, with the ordering being unknown to the decision maker in advance. We note that $\nu_i$ can be equal to $\nu_{i+1}$, i.e., one node can affect more than one of its neighbors simultaneously. Then $\nu_\eta$ is the first time when at least $\eta$ nodes are affected by the event. Thus, our problem is to detect the change at $\nu_\eta$ as quickly as possible subject to false alarm constraints.

For any $\boldsymbol{\nu}$, denote by $C(\boldsymbol{\nu}) = \{i : \nu_i < \infty\}$ the set of all the indices of the nodes that will eventually be affected by the event. Then

$$|C(\boldsymbol{\nu})| = \sum_{1 \leq i \leq L} \mathbb{1}_{\{\nu_i < \infty\}} \tag{2}$$

is the total number of affected nodes. If $|C(\boldsymbol{\nu})| = L$, then all the nodes will be affected by the event eventually.

We use $\mathbb{P}_{\boldsymbol{\nu}}$ to denote the probability measure of the samples with the set of change-points being $\boldsymbol{\nu}$, and let $\mathbb{E}_{\boldsymbol{\nu}}$ denote the corresponding expectation. For a given $\boldsymbol{\nu}$, if $|C(\boldsymbol{\nu})| < \eta$, i.e., $\nu_\eta = \infty$, then there are fewer than $\eta$ nodes that will be affected under $\mathbb{P}_{\boldsymbol{\nu}}$. In this case, if an alarm is triggered, it is a false alarm. For any stopping time $\tau$, to measure how frequently false alarms occur, we define the worst-case average run length (WARL) to false alarm as follows:

$$\text{WARL}(\tau) = \inf_{\boldsymbol{\nu}:|C(\boldsymbol{\nu})|<\eta} \mathbb{E}_{\boldsymbol{\nu}}[\tau]. \tag{3}$$

Then with a larger WARL, we have fewer false alarms.

Let $d_i = \nu_{i+1} - \nu_i$ denote the time it takes for the event to propagate from node $i$ to node $i+1$, for $1 \leq i \leq L-1$. If $d_i = 0$, then node $i$ and node $i+1$ are affected simultaneously. Denote $\boldsymbol{D} := \{d_\eta, d_{\eta+1}, \ldots, d_{L-1}\}$. For a fixed $\boldsymbol{D}$, to measure how quickly we can detect when at least $\eta$ nodes are affected, we define the worst-case average detection delay (WADD) using a criterion based on Pollak's criterion [37] as follows:

$$J_{\boldsymbol{D}}(\tau) = \sup_{\nu_1 \leq \cdots \leq \nu_\eta < \infty} \mathbb{E}_{\boldsymbol{\nu}}[\tau - \nu_\eta | \tau \geq \nu_\eta]. \tag{4}$$

We note that the supremum in (4) is only taken over $\nu_1 \leq \cdots \leq \nu_\eta < \infty$. Therefore, $J_{\boldsymbol{D}}[\tau]$ is a function of $\boldsymbol{D}$.

We denote by $\mathcal{F}_k$ the $\sigma$-algebra generated by the observations of all the nodes up to time $k$, for $k = 1, 2, \ldots$. We wish to find a $\{\mathcal{F}_k\}_{k \in \mathbb{N}}$-stopping time that achieves "small" detection delay, while controlling the false alarm rate. More specifically, for any $\boldsymbol{D}$, the goal is to minimize $J_{\boldsymbol{D}}[\tau]$ subject to a constraint on the WARL:

$$\inf_{\tau:\text{WARL}(\tau) \geq \gamma} J_{\boldsymbol{D}}(\tau). \tag{5}$$

To describe the objective in words, we want to find stopping rules so that for all possible scenarios with fewer than $\eta$ affected nodes, the average run length to false alarm is at least $\gamma$. At the same time, among those stopping rules that satisfy the false alarm requirement, we want to find the one that minimizes the WADD for all propagation dynamics after $\eta$ nodes are affected. There is no guarantee that the optimization problem in (5) has a solution, since we require the same stopping rule to simultaneously minimize the WADD for all propagation dynamics after $\eta$ nodes are affected. What we will show in the following sections is that such a "uniformly" optimum solution can be found up to a first-order approximation in an appropriately defined asymptotic setting.

**Notation**

We denote the samples across all the nodes at time $k$ by $\boldsymbol{X}[k] = \{X_1[k], \ldots, X_L[k]\}$, and the samples across all the nodes from time $k_1$ to $k_2$ by $\boldsymbol{X}[k_1, k_2] = \{\boldsymbol{X}[k_1], \ldots, \boldsymbol{X}[k_2]\}$. We further define

$$Z_i[k_1, k_2] = \sum_{k=k_1}^{k_2} \log \frac{f_1(X_i[k])}{f_0(X_i[k])}, \tag{6}$$

which is the log-likelihood ratio for the samples at node $i$ from time $k_1$ to $k_2$. We use the following conventions: $\sum_{j=k_1}^{k_2} A_j = 0$ and $\prod_{j=k_1}^{k_2} A_j = 1$ if $k_1 > k_2$. We use $X^+$ to denote the positive part of

$X$, i.e., $X^+ = \max\{X, 0\}$. We denote the Kullback-Leibler (KL) divergence between $f_1$ and $f_0$ as

$$I = \int f_1(x) \log \frac{f_1(x)}{f_0(x)} dx, \tag{7}$$

which is assumed to be positive and finite. We denote $x = o(1)$, as $c \to c_0$, if $\forall \epsilon > 0$, $\exists \delta > 0$, s.t., $|x| \leq \epsilon$ if $|c - c_0| < \delta$. We denote $g(c) \sim h(c)$, as $c \to c_0$, if $\lim_{c \to c_0} \frac{h(c)}{g(c)} = 1$.

## 3 Fully Connected Networks

In this section, we study fully connected networks, for which $G$ is a complete graph. In this case, the event can propagate from any node to any other node, and the induced sub-graph on any subset of nodes is connected. We present the Spartan-CuSum (S-CuSum) algorithm, and show that it can be implemented efficiently with complexity that is linear in $L$ at each time step. We then establish a lower bound on the WARL for the S-CuSum algorithm, and show how to choose the parameter to satisfy the false alarm constraint.

### 3.1 The S-CuSum Algorithm

We reformulate the quickest detection problem in Section 2 when $G$ is a complete graph as a dynamic composite hypothesis testing problem, i.e., to distinguish the following two hypotheses at each time $k$:

$$\mathcal{H}_0[k] : \sum_{i=1}^{L} \mathbb{1}_{\{\nu_i \leq k\}} < \eta, \tag{8}$$

$$\mathcal{H}_1[k] : \sum_{i=1}^{L} \mathbb{1}_{\{\nu_i \leq k\}} \geq \eta. \tag{9}$$

Both the null and alternative hypotheses are composite, since the data generating distribution depends on unknown $\boldsymbol{\nu}$ under each hypothesis. This hypothesis testing procedure stops once a decision in favor of the alternative hypothesis is reached; otherwise, a new sample is taken from each node in the network.

To distinguish between the two hypotheses, we consider the log-likelihood ratio between them. Since $\boldsymbol{\nu}$ under each hypothesis is unknown, we take a maximum likelihood approach with respect to the unknown $\boldsymbol{\nu}$, and construct the following generalized log-likelihood ratio statistic:

$$W[k] = \log \left( \frac{\max_{\boldsymbol{\nu}: \sum_{i=1}^{L} \mathbb{1}_{\{\nu_i \leq k\}} \geq \eta} \mathbb{P}_{\boldsymbol{\nu}}(\boldsymbol{X}[1,k])}{\max_{\boldsymbol{\nu}: \sum_{i=1}^{L} \mathbb{1}_{\{\nu_i \leq k\}} < \eta} \mathbb{P}_{\boldsymbol{\nu}}(\boldsymbol{X}[1,k])} \right). \tag{10}$$

The max in the numerator in (10) is taken over all $\boldsymbol{\nu}$ such that $\sum_{i=1}^{L} \mathbb{1}_{\{\nu_i \leq k\}} \geq \eta$, where at time $k$, there are no fewer than $\eta$ affected nodes. Likewise, the max in the denominator in (10) is taken over all $\boldsymbol{\nu}$ such that $\sum_{i=1}^{L} \mathbb{1}_{\{\nu_i \leq k\}} < \eta$, where at time $k$, there are fewer than $\eta$ affected nodes.

Since the network is fully connected, the induced sub-graph on any subset of nodes is connected. Therefore, the max in both the numerator and denominator in (10) is taken without explicitly enforcing connectivity and propagation dynamics.

The corresponding stopping time is then given by comparing $W[k]$ against a pre-determined positive threshold:

$$\tilde{\tau}(b) = \inf\{k \geq 1 : W[k] > b\}, \tag{11}$$

where $b$ will be selected according to the false alarm constraint.

## 3.2 A Simpler but Equivalent Form

In this subsection, we develop an equivalent but much simpler form of (11), which can be computed with complexity $O(L)$ at each time step.

Let $\mathbb{P}_\infty$ denote the probability measure with $\nu_i = \infty, \forall 1 \leq i \leq L$. Then, it can be easily shown that

$$W[k] = \max_{\boldsymbol{\nu}:\sum_{i=1}^{L} \mathbb{1}_{\{\nu_i \leq k\}} \geq \eta} \log\left(\frac{\mathbb{P}_{\boldsymbol{\nu}}(\boldsymbol{X}[1,k])}{\mathbb{P}_\infty(\boldsymbol{X}[1,k])}\right)$$
$$- \max_{\boldsymbol{\nu}:\sum_{i=1}^{L} \mathbb{1}_{\{\nu_i \leq k\}} < \eta} \log\left(\frac{\mathbb{P}_{\boldsymbol{\nu}}(\boldsymbol{X}[1,k])}{\mathbb{P}_\infty(\boldsymbol{X}[1,k])}\right). \tag{12}$$

Due to the fact that

$$\log\left(\frac{\mathbb{P}_{\boldsymbol{\nu}}(\boldsymbol{X}[1,k])}{\mathbb{P}_\infty(\boldsymbol{X}[1,k])}\right)$$
$$= \log\left(\prod_{i=1}^{L} \frac{\prod_{j=1}^{\min\{\nu_i-1,k\}} f_0(X_i[j]) \prod_{j=\nu_i}^{k} f_1(X_i[j])}{\prod_{j=1}^{k} f_0(X_i[j])}\right)$$
$$= \sum_{i=1}^{L} \sum_{j=\nu_i}^{k} \log \frac{f_1(X_i[j])}{f_0(X_i[j])}, \tag{13}$$

the first term in (12) is equivalent to

$$\max_{\boldsymbol{\nu}:\sum_{i=1}^{L} \mathbb{1}_{\{\nu_i \leq k\}} \geq \eta} \sum_{i=1}^{L} \sum_{j=\nu_i}^{k} \log \frac{f_1(X_i[j])}{f_0(X_i[j])}. \tag{14}$$

Similarly, the second term in (12) is equivalent to

$$\max_{\boldsymbol{\nu}:\sum_{i=1}^{L} \mathbb{1}_{\{\nu_i \leq k\}} < \eta} \sum_{i=1}^{L} \sum_{j=\nu_i}^{k} \log \frac{f_1(X_i[j])}{f_0(X_i[j])}. \tag{15}$$

If we denote the individual CuSum statistic [35] at node $i$ (testing a change from $f_0$ to $f_1$) at time $k$ as

$$W_i[k] = \max_{1 \leq \nu_i \leq k} \sum_{j=\nu_i}^{k} \log \frac{f_1(X_i[j])}{f_0(X_i[j])}, \tag{16}$$

and define a permutation $\mu(\cdot)$ such that

$$W_{\mu(1)}[k] \geq W_{\mu(2)}[k] \geq \cdots \geq W_{\mu(L)}[k], \tag{17}$$

then, $\tilde{\tau}(b)$ is equivalent to

$$\hat{\tau}(b) = \inf\left\{k \geq 1 : \sum_{i=\eta}^{L} \left(W_{\mu(i)}[k]\right)^+ \geq b\right\}, \tag{18}$$

which we refer to as the S-CuSum algorithm. Such an equivalence can be established as follows.

1. If $W_{\mu(\eta)}[k] \geq 0$, then (14) is equal to $\sum_{i=1}^{L} \left(W_{\mu(i)}[k]\right)^+$, and (15) is equal to $\sum_{i=1}^{\eta-1} W_{\mu(i)}[k]$. It then follows that $W[k] = \sum_{i=\eta}^{L} \left(W_{\mu(i)}[k]\right)^+$.

2. If $W_{\mu(\eta)}[k] < 0$, then (14) is equal to $\sum_{i=1}^{\eta} W_{\mu(i)}[k]$, and (15) is equal to $\sum_{i=1}^{\eta-1} \left(W_{\mu(i)}[k]\right)^+$. In this case, $W[k]$ is non-positive, and $\sum_{i=\eta}^{L} \left(W_{\mu(i)}[k]\right)^+ = 0$. Since $b$ is positive, the test in (11) is equivalent to comparing $\sum_{i=\eta}^{L} \left(W_{\mu(i)}[k]\right)^+$ to $b$.

The test in (18) can be implemented efficiently. First of all, for each node $i$, $W_i[k]$ can be updated recursively:

$$W_i[k] = (W_i[k-1])^+ + \log \frac{f_1(X_i[k])}{f_0(X_i[k])}. \tag{19}$$

Second, we do not need to sort all $W_i[k]$ at each time $k$. We only need to find the smallest $L-\eta+1$ numbers from $L$ numbers, which can be solved with $O(L)$ computational cost using the algorithm in [38] instead of $O(L \log L)$. Thus, the total computational cost at each time $k$ is $O(L)$. Here, we note that, at time $k$, each node $i$ may choose to send $X_i[k]$, $\log \frac{f_1(X_i[k])}{f_0(X_i[k])}$, or $W_i[k]$ to the fusion center.

### 3.3 Lower Bound on the WARL

The following theorem provides a lower bound on the WARL for the S-CuSum algorithm.

**Theorem 1.** *The WARL for the S-CuSum algorithm in (18) is lower bounded as follows:*

$$WARL(\hat{\tau}(b)) \geq \frac{1}{poly(b)} e^b, \tag{20}$$

*where poly(b) denotes a polynomial of $b$.*

*Proof.* To show the lower bound on WARL($\hat{\tau}(b)$), it suffices to show that for any $\boldsymbol{\nu}$ with $\nu_i = \infty$, $\forall \eta \leq i \leq L$,

$$\mathbb{E}_{\boldsymbol{\nu}}[\hat{\tau}(b)] \geq \frac{1}{\text{poly}(b)} e^b. \tag{21}$$

For any $t \in \mathbb{N}$ and $b > 0$, it follows that

$$\mathbb{P}_{\boldsymbol{\nu}}(\hat{\tau}(b) \leq t)$$
$$= \mathbb{P}_{\boldsymbol{\nu}}\left(\max_{1 \leq k \leq t} \sum_{i=\eta}^{L} \left(W_{\mu(i)}[k]\right)^+ > b\right)$$
$$\leq \sum_{k=1}^{t} \mathbb{P}_{\boldsymbol{\nu}}\left(\sum_{i=\eta}^{L} \left(W_{\mu(i)}[k]\right)^+ > b\right). \tag{22}$$

Since $\sum_{i=\eta}^{L} \left(W_{\mu(i)}[k]\right)^+ \leq \sum_{i=\eta}^{L} \left(W_i[k]\right)^+$, we have

$$\mathbb{P}_{\boldsymbol{\nu}} \left( \sum_{i=\eta}^{L} \left(W_{\mu(i)}[k]\right)^+ > b \right)$$
$$\leq \mathbb{P}_{\boldsymbol{\nu}} \left( \sum_{i=\eta}^{L} \left(W_i[k]\right)^+ > b \right). \tag{23}$$

By [13, Lemma B1], it then follows that

$$\mathbb{P}_{\boldsymbol{\nu}} \left( \sum_{i=\eta}^{L} \left(W_i[k]\right)^+ > b \right) \leq \text{poly}(b) e^{-b}, \tag{24}$$

where $\text{poly}(b)$ has an order of $L - \eta$. Therefore, $\mathbb{P}_{\boldsymbol{\nu}}(\hat{\tau}(b) \leq t) \leq t \cdot \text{poly}(b) e^{-b}$, which implies that

$$\begin{aligned}
\mathbb{E}_{\boldsymbol{\nu}}[\hat{\tau}(b)] &= \sum_{t=0}^{\infty} \mathbb{P}_{\boldsymbol{\nu}}(\hat{\tau}(b) \geq t) \\
&\geq \sum_{t=0}^{\infty} (1 - t \cdot \text{poly}(b) e^{-b})^+ \\
&= \sum_{t=0}^{e^b/\text{poly}(b)} \left(1 - t \cdot \text{poly}(b) e^{-b}\right) \\
&= \frac{1}{\text{poly}(b)} e^b.
\end{aligned} \tag{25}$$

□

**Corollary 1.** *To guarantee $WARL(\hat{\tau}(b)) \geq \gamma$, it suffices to choose $b$ such that*

$$\frac{1}{poly(b)} e^b = \gamma, \tag{26}$$

*and $b \sim \log \gamma$, as $\gamma \to \infty$.*

*Proof.* The result follows from Theorem 1. □

## 4 Asymptotic Analysis

In this section we study the asymptotic performance of the proposed S-CuSum algorithm in (18) and demonstrate its asymptotic optimality. For our asymptotic analysis to be non-trivial, we let not only the prescribed lower bound on the WARL, $\gamma$, go to infinity, but also $d_\eta, d_{\eta+1}, \ldots, d_{L-1}$. Indeed, if the latter variables are fixed as $\gamma$ goes to infinity, then we will not be able to characterize how the propagation dynamics affects the performance, and the asymptotic performance will only depend on the number of nodes that will be affected eventually. Therefore, in order to perform a general and relevant asymptotic analysis, we let $d_\eta, d_{\eta+1}, \ldots, d_{L-1}$ go to infinity with $\gamma$. Without loss of generality, suppose that

$$d_{\eta+i-1} \sim c_i \frac{\log \gamma}{iI}, \qquad (27)$$

as $\gamma \to \infty$, where $c_i \in [0, \infty]$ for $i = 1, \ldots, L - \eta$ are unknown. We further assume that $d_L = \infty$, $c_{L-\eta+1} = \infty$. Here if $d_{\eta+i-1}$ has an order less than $\log \gamma$, then $c_i = 0$, and if $d_{\eta+i-1}$ has an order greater than $\log \gamma$, then $c_i = \infty$. Such an asymptotic setting is only used for the convenience of analysis, and as an approximation of the performance when $d_\eta, d_{\eta+1}, \ldots, d_{L-1}$ and $\gamma$ are large. Applying the algorithm in practice does not rely on this assumption.

In the following, we first present an example with $L = 3$ and $\eta = 2$ to understand the results, and then move on to the general results.

## 4.1 Example: $L = 3$ and $\eta = 2$

Consider a fully connected network with three nodes, i.e., $L = 3$. Our goal is to detect when at least two nodes are affected, i.e., $\eta = 2$. Then the S-CuSum algorithm is equivalent to comparing the sum of the smallest two individual CuSum statistics to a threshold $b$:

$$\hat{\tau}(b) = \inf \left\{ k \geq 1 : \min_{1 \leq i < j \leq 3} (W_i[k])^+ + (W_j[k])^+ \geq b \right\}. \qquad (28)$$

For simplicity, we use the notion of phase $i$ to denote the phase in which there are $i$ affected nodes. Then after an event occurs, the network first changes from phase 0 to phase 1, then to phase 2, and eventually stabilizes in phase 3 (see Fig. 2).

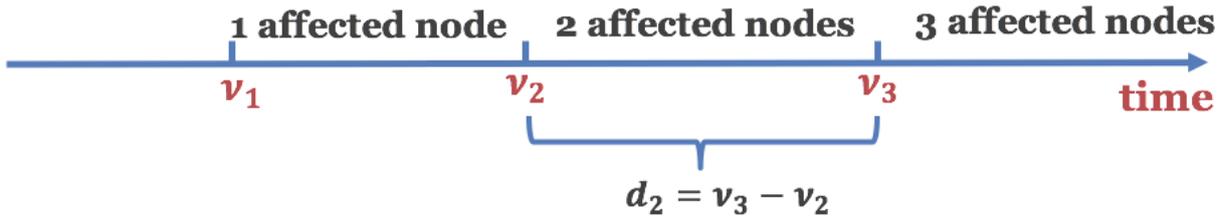

Figure 2: Example with $L = 3$.

For this example, the (first-order) asymptotic optimality of the S-CuSum algorithm is characterized in the following proposition.

**Proposition 1.** *Let the threshold $b \sim \log \gamma$ so that $WARL(\hat{\tau}(b)) \geq \gamma$. Assume that $d_2$ and $\gamma$ go to infinity as in (27), then the S-CuSum is asymptotically optimal:*

$$J_{\boldsymbol{D}}(\hat{\tau}(b)) \sim \inf_{\tau: \text{WARL}(\tau) \geq \gamma} J_{\boldsymbol{D}}(\tau)$$

$$\sim \begin{cases} \dfrac{\log \gamma}{I}, & \text{if } c_1 > 1, \\ \log \gamma \left( \dfrac{c_1}{I} + \dfrac{1 - c_1}{2I} \right), & \text{if } c_1 \leq 1. \end{cases} \qquad (29)$$

*Proof.* The proof for this special case is omitted. See detailed proof for the general case in Theorem 4. □

The optimal performance shows a dichotomy depending on whether $c_1 > 1$ or not. In the following, we will first provide a heuristic explanation for the dichotomy, and then provide the results for the general case together with rigorous proofs.

Roughly speaking, if we consider the CuSum statistic at node $i$, before the change-point $\nu_i$, it is small and close to 0, and after the change-point, it grows with a positive slope of $I$. Therefore, by (28), the S-CuSum is close to 0 in phases 0 and 1, and grows with slope $I$ in phase 2 and with slope $2I$ in phase 3.

If $d_2 I > b$, i.e., $c_1 > 1$, then the S-CuSum statistic crosses the threshold $b$ within phase 2 (see Fig. 3). The detection delay for this case is $b/I$. If $d_2 I \leq b$, i.e., $c_1 \leq 1$, then the S-CuSum statistic is not large enough to cross the threshold $b$ within phase 2, and it needs more samples from phase 3 (see Fig. 4). The detection delay is then equal to the sum of the duration of phase 2 and the number of samples needed from phase 3:

$$d_2 + \frac{b - d_2 I}{2I} \sim b\left(\frac{c_1}{I} + \frac{1 - c_1}{2I}\right). \tag{30}$$

Depending on whether or not phase 2 is long enough, the performance of the S-CuSum algorithm shows a dichotomy.

## 4.2 Asymptotic Universal Lower Bound on the WADD

In this subsection, we study the universal lower bound on the WADD for any stopping rule with the WARL no smaller than $\gamma$. We denote

$$h = \inf\{1 \leq j \leq L - \eta + 1 : \sum_{i=1}^{j} c_i \geq 1\}. \tag{31}$$

**Theorem 2.** *Suppose* (27) *holds. Then as* $\gamma \to \infty$,

$$\inf_{\tau: \text{WARL}(\tau) \geq \gamma} J_{\boldsymbol{D}}(\tau)$$

$$\geq \log \gamma \left(\sum_{i=1}^{h-1} \frac{c_i}{iI} + \frac{1 - \sum_{i=1}^{h-1} c_i}{hI}\right)(1 - o(1)). \tag{32}$$

*Proof.* See Appendix B. □

Theorem 2 suggests that to meet the asymptotic universal lower bound, an algorithm should be adaptive to the unknown $d_\eta, d_{\eta+1}, \ldots, d_{L-1}$. An intuitive understanding of $h$ is that the algorithm shall stop within phase $h + \eta$, when there are $h + \eta$ affected nodes.

The proof is based on a change-of-measure argument and a Law of Large Numbers argument for the log-likelihood ratio statistics, similar to those in [22]. However, a major difference in the change-of-measure argument compared to [22] is that the "pre-change" mode is composite, i.e., there are multiple possible scenarios with fewer than $\eta$ affected nodes. Furthermore, the post-change statistic is more complicated, since the propagation dynamics is unknown, and the number of affected nodes is changing with time.

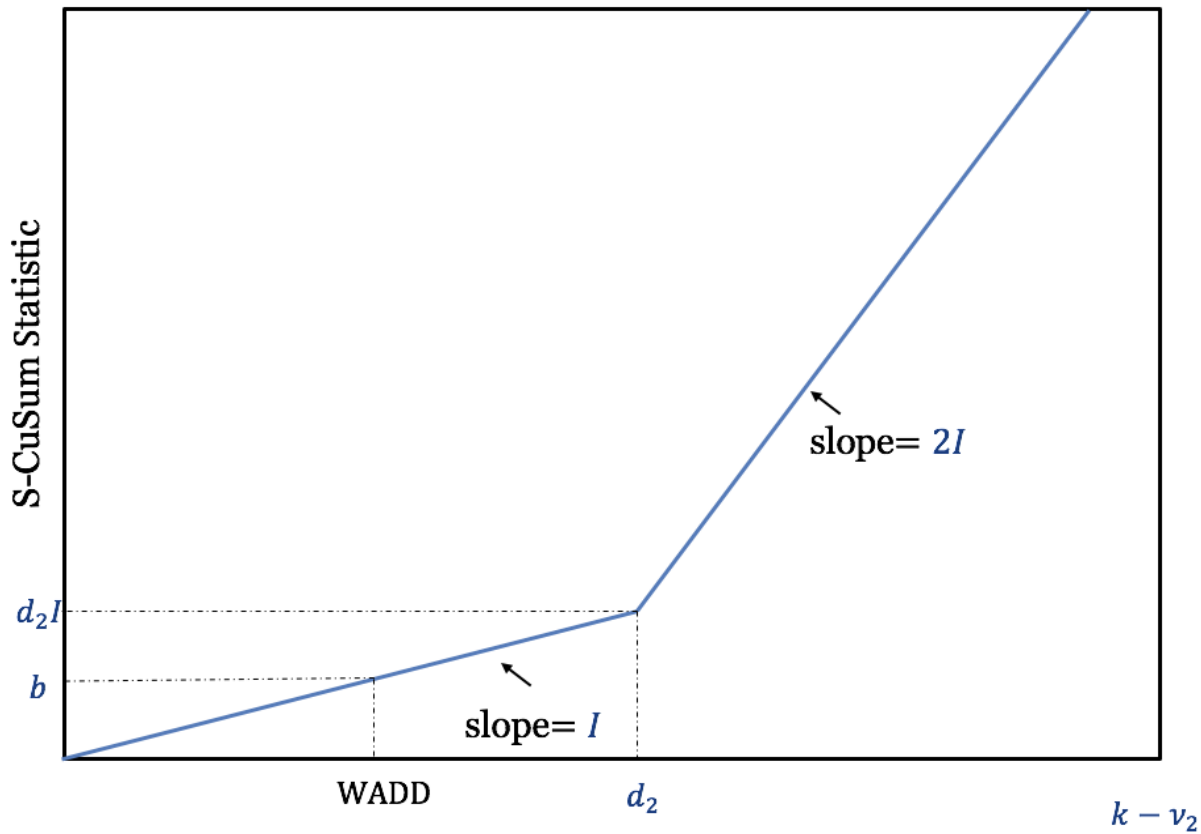

Figure 3: Scenario 1: $d_2$ is large.

### 4.3 Asymptotic Upper Bound on the WADD

Recall that we can choose $b \sim \log \gamma$ such that the false alarm constraint is satisfied. Then, by (27), it follows that

$$d_{\eta+i-1} \sim c_i \frac{b}{iI}, \tag{33}$$

as $\gamma \to \infty$, where $c_i \in [0, \infty]$, for every $i = 1, \ldots, L - \eta$.

An asymptotic upper bound on the WADD for the S-CuSum algorithm in (18) is characterized in the following theorem.

**Theorem 3.** *Suppose (33) holds. Then as $b \to \infty$,*

$$J_{\boldsymbol{D}}(\hat{\tau}(b)) \leq b \left( \sum_{i=1}^{h-1} \frac{c_i}{iI} + \frac{1 - \sum_{i=1}^{h-1} c_i}{hI} \right) (1 + o(1)). \tag{34}$$

*Proof.* See Appendix C. □

From Theorem 3, it is clear that although the S-CuSum algorithm does not exploit the knowledge of $d_\eta, d_{\eta+1}, \ldots, d_{L-1}$, the performance is still adaptive to the unknown $d_\eta, d_{\eta+1}, \ldots, d_{L-1}$. This is consistent with the insights from the asymptotic universal lower bound in Theorem 2.

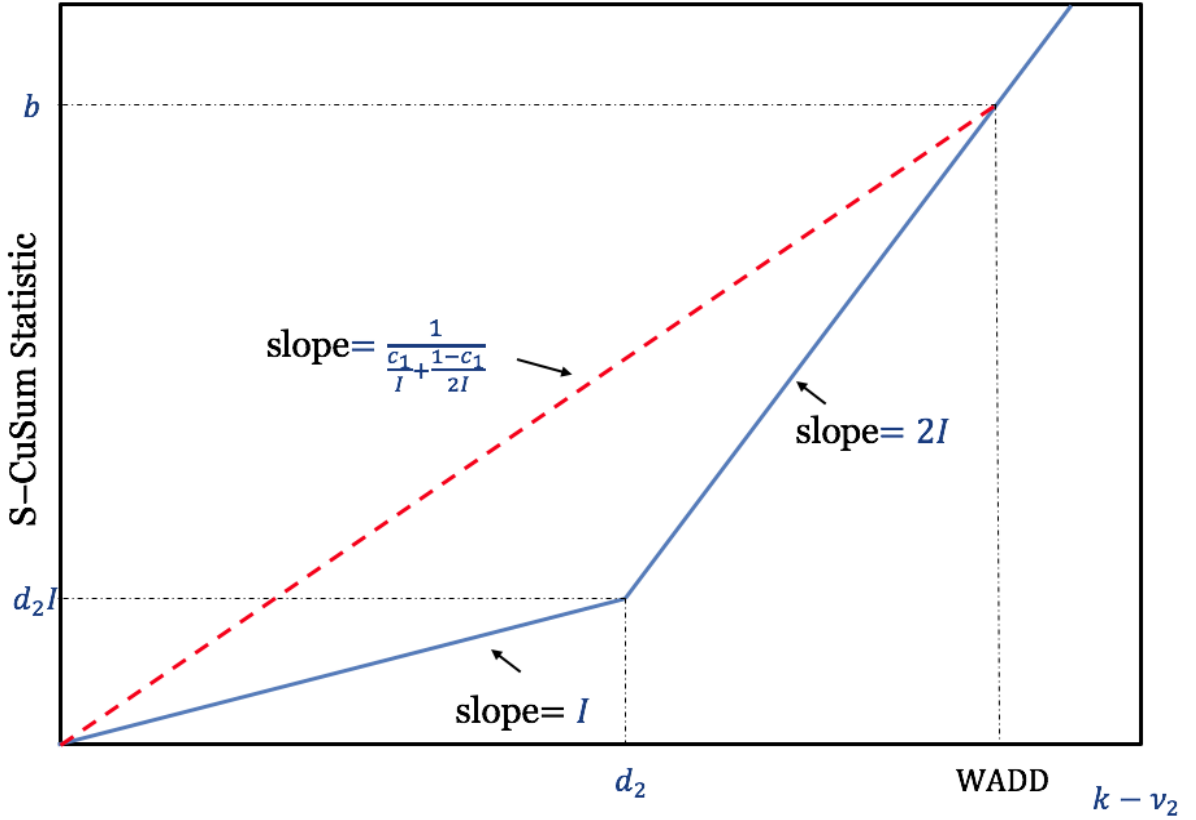

Figure 4: Scenario 2: $d_2$ is small.

The proof of the asymptotic upper bound on WADD is based on partitioning the samples into independent blocks and applying the Law of Large Numbers for the log-likelihood ratio statistics, as in [22, Theorem 4]. The major difficulty here is due to the more complicated test statistic, in which the number of affected nodes changes with time.

### 4.4 Asymptotic Optimality of S-CuSum

We are now ready to establish the asymptotic optimality of the S-CuSum algorithm, which is presented in the following theorem.

**Theorem 4** (S-CuSum, Asymptotic Optimality). *Let the threshold $b \sim \log \gamma$ so that $WARL(\hat{\tau}(b)) \geq \gamma$. Assume that $d_\eta, d_{\eta+1}, \ldots, d_{L-1}$ and $\gamma$ go to infinity as in (27), then the S-CuSum algorithm is asymptotically optimal:*

$$J_{\boldsymbol{D}}(\hat{\tau}(b)) \sim \inf_{\tau: \mathrm{WARL}(\tau) \geq \gamma} J_{\boldsymbol{D}}(\tau)$$
$$\sim \log \gamma \left( \sum_{i=1}^{h-1} \frac{c_i}{iI} + \frac{1 - \sum_{i=1}^{h-1} c_i}{hI} \right). \qquad (35)$$

*Proof.* This result follows from Theorems 1, 2 and 3. □

# 5 Arbitrarily Connected Networks

In this section, we extend results of the previous section to arbitrarily connected networks.

Since we assume that the event propagates along edges in the network, the induced sub-graph on the affected nodes is then connected at each time step. Here, for an arbitrarily connected network, the induced sub-graph on some subset of nodes may not be connected. Therefore, although the S-CuSum algorithm still applies, and it is asymptotically optimal (up to a first-order approximation), it will have more false alarms due to the fact that it may raise alarms when it detects $\eta$ nodes that are not connected (as we will show numerically in Section 6).

To exploit knowledge of the network structure, one can directly adapt the generalized log-likelihood ratio test in (10). However, this will involve a scan statistic over all possible propagation dynamics along the edges in the network, which leads to a combinatorial problem over a large search space. Therefore, it is computationally infeasible, especially for large networks.

In the following, we present the Network-CuSum (N-CuSum) algorithm, which not only employs knowledge of the network structure, but does so in a computationally efficient way. We then establish the first-order asymptotic optimality of the N-CuSum algorithm. Also, as will be shown in the numerical results in Section 6, for an arbitrarily connected network, the N-CuSum algorithm performs much better than the S-CuSum.

## 5.1 The N-CuSum Algorithm

At each time step $k$, we update the local CuSum statistics $W_i[k]$, $\forall 1 \leq i \leq L$. We then compare each local CuSum statistic to a threshold $\log b$, and delete this node if its CuSum statistic is less than $\log b$. The resulting graph is then denoted by $G'[k]$. For this step, the computational complexity is $O(L)$. These deleted nodes are highly likely to be not affected by the event.

We run the BFS algorithm on $G'[k]$ to recover all connected components of $G'[k]$: $C_1[k], C_2[k], \ldots$. The computational complexity for this step is at most $O(L + |E|)$. We then run the S-CuSum algorithm on each connected component, and use S-CuSum$_i[k]$ to denote the test statistic value of the S-CuSum algorithm on $C_i[k]$:

$$\text{S-CuSum}_i[k] = \min_{\substack{C' \subseteq C_i[k]: \\ |C'| = |C_i[k]| - \eta + 1}} \sum_{i \in C'} (W_i[k])^+ . \qquad (36)$$

If any of these statistics crosses the threshold $b$, we stop and raise an alarm. For this step, the computational complexity is less than $O(L)$. Therefore, the overall computational complexity at each time step is $O(L + |E|)$, which means that this algorithm scales well as the network size grows, assuming that the network is not dense. The N-CuSum algorithm is described in detail in Algorithm 1.

## 5.2 Performance Analysis of the N-CuSum Algorithm

We next provide theoretical analysis for the N-CuSum algorithm. The following theorem provides a lower bound on the WARL for the N-CuSum algorithm.

**Algorithm 1** N-CuSum

**Input:**
$G$: graph
$\eta$: size of sub-graph of interest
$f_0, f_1$: distributions before and after change
$b$: threshold
**Output:**
$\bar{\tau}$: stopping time
**Initialization:**
$W_i[0] \leftarrow 0$, for $1 \leq i \leq L$
$k = 0$
**Method:**
**while** 1 **do**
    $k \leftarrow k + 1$
    Observe $X_i[k]$, for $1 \leq i \leq L$
    $W_i[k] \leftarrow (W_i[k-1])^+ + \log \frac{f_1(X_i[k])}{f_0(X_i[k])}$
    $G'[k] \leftarrow G$
    **for** $i = 1$ to $L$ **do**
        **if** $W_i[k] \leq \log b$ **then**
            Delete node $i$ and all edges connected to node $i$ in $G'[k]$
        **end if**
    **end for**
    All connected components of $G'[k]$: $C_1[k], C_2[k], \ldots \leftarrow$ run BFS on $G'[k]$
    **for** i=1,2,... **do**
        S-CuSum$_i[k] \leftarrow$Run S-CuSum on $C_i[k]$
        **if** S-CuSum$_i[k] \geq b$ **then**
            $\bar{\tau} \leftarrow k$
            Break
        **end if**
    **end for**
**end while**
Return $\bar{\tau}$

**Theorem 5.** *The WARL for the N-CuSum algorithm is lower bounded as follows:*

$$WARL(\bar{\tau}(b)) \geq \frac{1}{poly(b)} e^b. \tag{37}$$

*Proof.* It can be shown that S-CuSum$_i[k]$ is less than the S-CuSum statistic applied on the whole network, for any $i$ and $k$. Therefore,

$$\text{WARL}(\bar{\tau}(b)) \geq \text{WARL}(\hat{\tau}(b)). \tag{38}$$

Together with Theorem 1, this completes the proof. □

To guarantee WARL$(\bar{\tau}(b)) \geq \gamma$, it suffices to choose $b$ such that

$$\frac{1}{\text{poly}(b)} e^b = \gamma, \tag{39}$$

and $b \sim \log \gamma$.

For the asymptotic analysis, we choose the same asymptotic setting as in (27) in Section 4. By choosing $b \sim \log \gamma$, we also have (33). We then have the asymptotic upper bound on the WADD for the N-CuSum algorithm as characterized in the following theorem.

**Theorem 6.** *Suppose* (33) *holds. Then as* $b \to \infty$,

$$J_{\boldsymbol{D}}(\bar{\tau}(b)) \leq b \left( \sum_{i=1}^{h-1} \frac{c_i}{iI} + \frac{1 - \sum_{i=1}^{h-1} c_i}{hI} \right) (1 + o(1)). \tag{40}$$

*Proof.* See Appendix D. □

The proof of the asymptotic upper bound is similar to that of Theorem 3, but requires a more careful construction. This is due to the dependency between the individual S-CuSum statistics and the partition of the graph $G'[k]$ into connected components. The asymptotic universal lower bound in Theorem 2 also applies to the arbitrarily connected network here. We then establish the asymptotic optimality of N-CuSum in the following theorem.

**Theorem 7** (N-CuSum, Asymptotic Optimality). *Let threshold* $b \sim \log \gamma$ *so that* $\mathrm{WARL}(\bar{\tau}(b)) \geq \gamma$. *Assume that* $d_\eta, d_{\eta+1}, \ldots, d_{L-1}$ *and* $\gamma$ *go to infinity as in* (27), *then the N-CuSum algorithm is asymptotically optimal:*

$$\begin{aligned} J_{\boldsymbol{D}}(\bar{\tau}(b)) &\sim \inf_{\tau: \mathrm{WARL}(\tau) \geq \gamma} J_{\boldsymbol{D}}(\tau) \\ &\sim \log \gamma \left( \sum_{i=1}^{h-1} \frac{c_i}{iI} + \frac{1 - \sum_{i=1}^{h-1} c_i}{hI} \right). \end{aligned} \tag{41}$$

*Proof.* This result follows from Theorems 2, 5 and 6. □

## 6 Numerical Results

In this section, we present some numerical results. We start with an example to demonstrate a typical evolution path of the S-CuSum algorithm. We then study a fully connected network with three nodes. We compare the S-CuSum algorithm to a generalization of the multi-chart CuSum algorithm in [36] which stops when at least $\eta$ local CuSums have crossed their individual thresholds. Finally, we study a network that is not fully connected, a lattice network with 36 nodes. In this example, we also consider a network generalized multi-chart CuSum algorithm, which is to wait until $\eta$ local CuSums have crossed their individual thresholds simultaneously, and those nodes form a connected subgraph. We compare the generalized multi-chart CuSum, the network generalized multi-chart CuSum, the S-CuSum and the N-CuSum algorithms. For all four algorithms, the communication complexity at each time step is $L$, since we are considering a centralized setting.

In Fig. 5, we plot the evolution paths of the S-CuSum statistic and all the individual CuSum statistics. We consider a fully connected network with $L = 3$ and $\eta = 2$. We choose $f_0 = \mathcal{N}(0,1)$, and $f_1 = \mathcal{N}(1,1)$. We set $\boldsymbol{\nu} = \{1, 40, 80\}$. There are in total three phases, depending on the number of affected nodes.

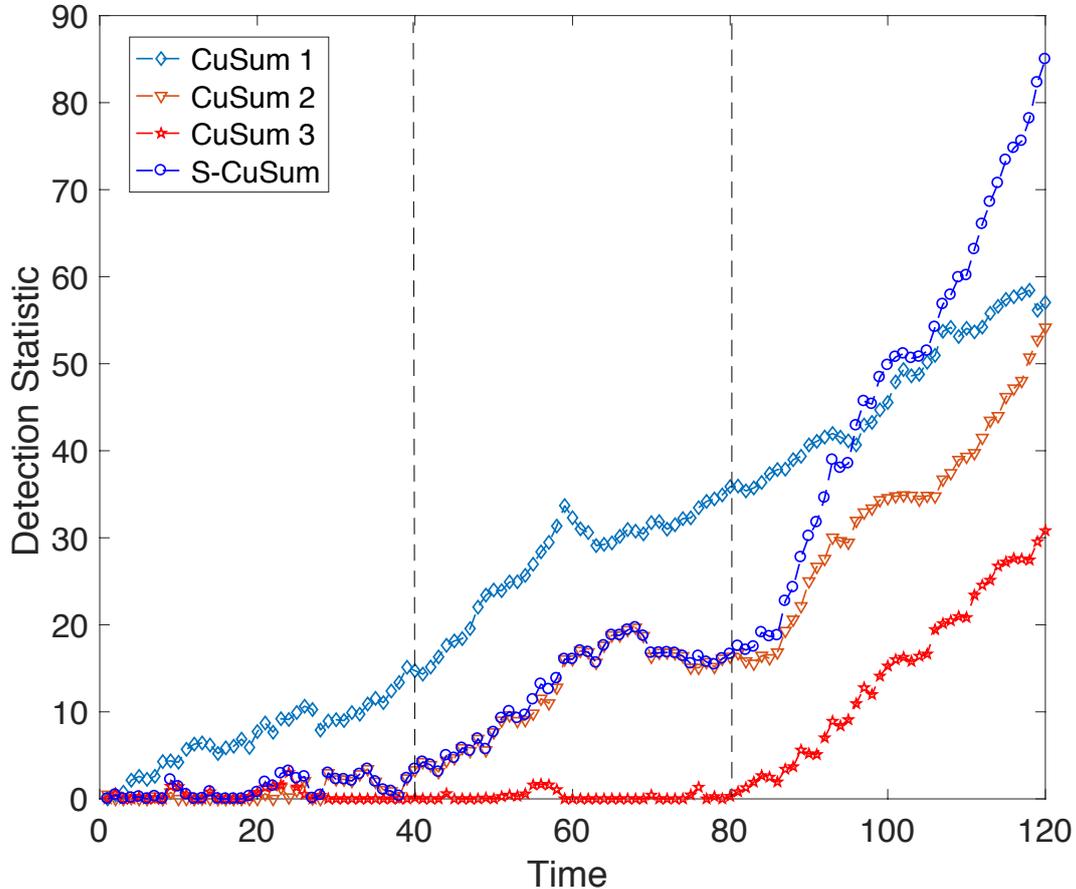

Figure 5: Sample evolution paths of all individual CuSums and the S-CuSum. CuSum $i$ denotes the individual CuSum statistic at node $i$, for $i = 1, 2, 3$.

In phase 1, i.e., $k < 40$, only the statistic of CuSum 1 grows with a positive slope, and all the other statistics are small and close to zero. Then, in phase 2, i.e., $40 \leq k < 80$, the statistics of CuSum 1, CuSum 2 and S-CuSum grow with a positive slope. In this phase, the S-CuSum statistic is almost the same as the CuSum 2 statistic, which is due to the fact that the S-CuSum statistic is the sum of the smallest two individual CuSum statistics, i.e. CuSum 2 and CuSum 3. Since the CuSum 3 statistic is small and close to zero, the S-CuSum statistic is almost the same as the CuSum 2 statistic in this phase. Eventually, in phase 3, i.e., $k \geq 80$, the statistics of CuSum 1, CuSum 2 and CuSum 3 all increase with a positive slope. In this phase, the S-CuSum statistic is the sum of the CuSum 2 and CuSum3 statistics. Therefore, the slope of the S-CuSum statistic is larger than that in phase 2.

In summary, the S-CuSum statistic is small and close to zero with only one affected node, and gradually grows but with different slopes with two and three affected nodes. Thus, the S-CuSum algorithm is adaptive to the unknown propagation dynamics.

We then study the performance of the S-CuSum algorithm, validate our theoretical assertions, and compare it with a generalization of the multi-chart CuSum algorithm in [36], which stops when at least $\eta$ local CuSum algorithms have crossed their individual thresholds simultaneously. Here also we consider a fully connected network with $L = 3$ and $\eta = 2$. We choose $f_0 = \mathcal{N}(0, 1)$, and $f_1 = \mathcal{N}(0.4, 1)$. Here we choose $\nu_1 = \nu_2 = 1$, and $d_2 = 40$ to simulate the average detection delay, and choose $\nu_1 = 1$, $\nu_2 = \nu_3 = \infty$ to simulate the average run length to false alarm. The plot is averaged over 1000 runs. We plot WADD versus WARL for the S-CuSum algorithm and the generalized multi-chart CuSum algorithm in Fig. 6.

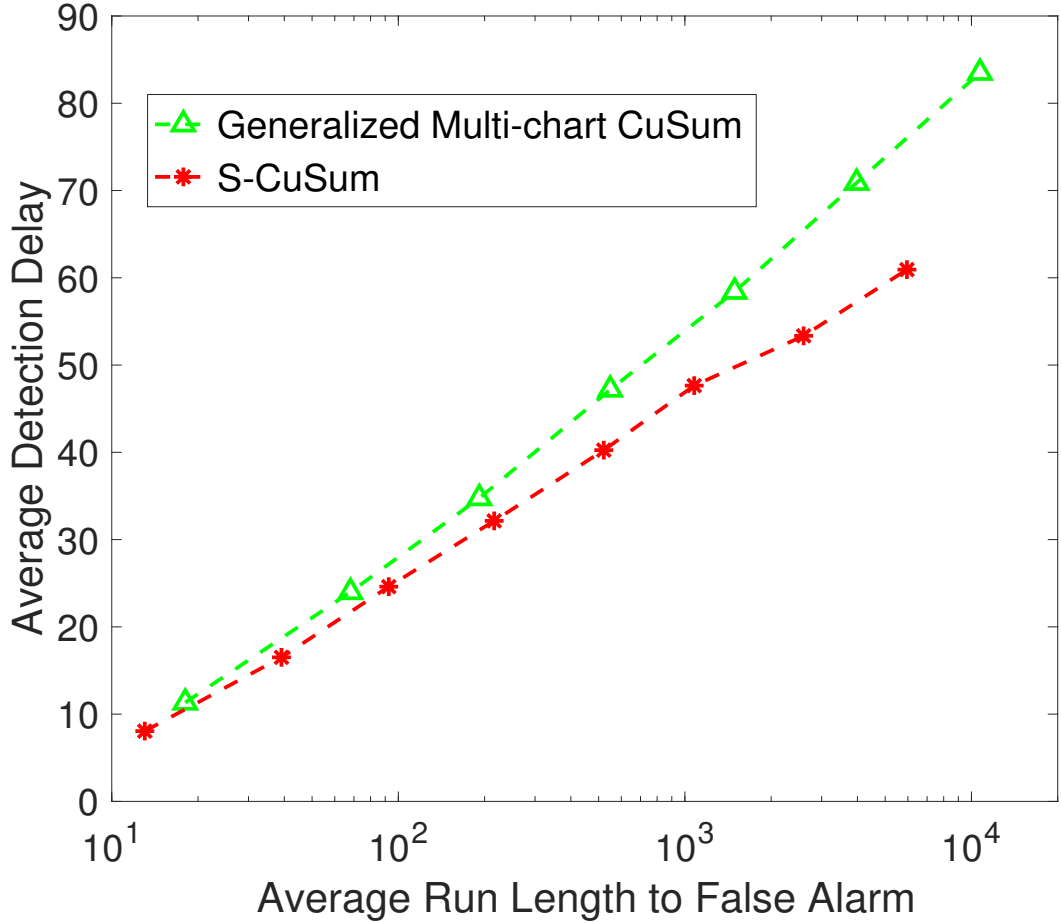

Figure 6: Comparison between the S-CuSum algorithm and the generalized multi-chart CuSum algorithm for a fully connected network.

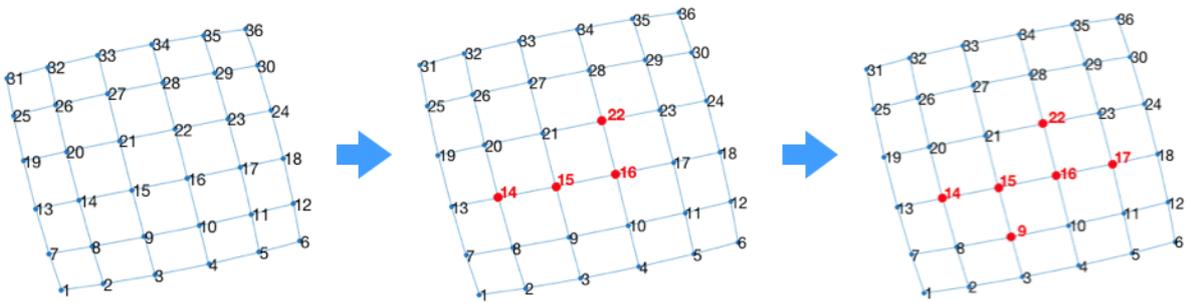

Figure 7: A dynamic event propagates in a lattice network.

In Fig. 6, the slope of the curve corresponding to the S-CuSum algorithm gradually changes after WADD= 40, which validates our theoretical results in Proposition 1. Furthermore, the S-CuSum algorithm performs better than the generalized multi-chart CuSum algorithm, especially when WADD$\geq d_2$. This is because when WADD$\geq d_2$, there are three affected nodes. The samples from the third affected node also contain information about whether there are no fewer than $\eta$ affected nodes (although the local CuSum statistic at the third affected node is not large), and this information is used by the S-CuSum algorithm but not the multichart CuSum algorithm.

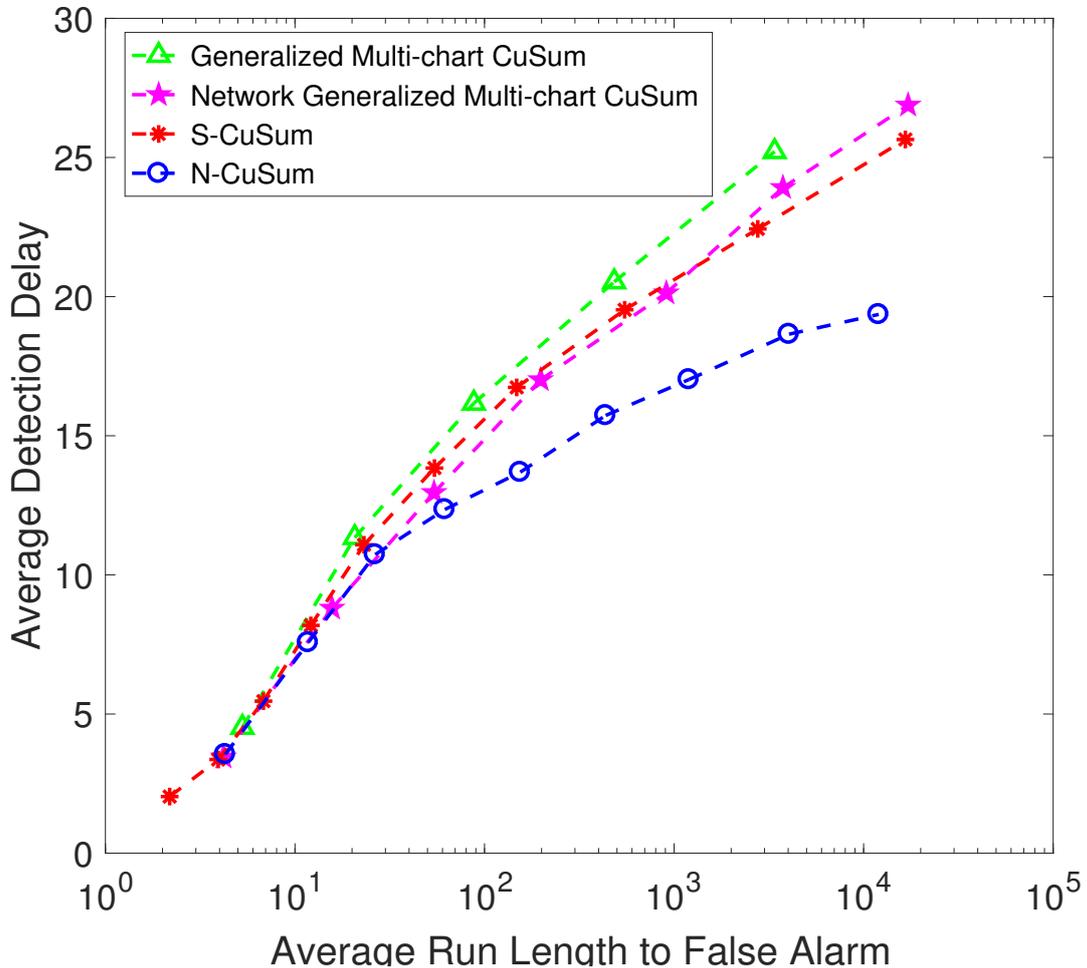

Figure 8: Comparison among the generalized multi-chart CuSum algorithm, the N-generalized multi-chart CuSum algorithm, the S-CuSum algorithm and the N-CuSum algorithm.

We next consider a lattice network with 36 nodes (see Fig. 7). We set $\eta = 4$, $f_0 = \mathcal{N}(0, 1)$, and $f_1 = \mathcal{N}(1, 1)$. To simulate the average detection delay, we assume that the event first affects nodes $14, 15, 16, 22$ at time 1, then propagates to nodes 9 and 17 at time 10, and no other node is affected by the event. To simulate the average run length to false alarm, we assume that nodes 14, 15, 16 are affected at time 1, and no other node is affected by the event. We repeat the simulation for 1000 times. As we can see from Fig. 8, the N-CuSum algorithm has the best performance among the four algorithms. Compared to the S-CuSum algorithm, the N-CuSum algorithm has significantly reduced the WADD, which is due to its effective exploitation of the network structure.

Under the same setting used to simulate the average detection delay as shown in Fig. 8, we plot the average number of connected components in $G'[k]$ when N-CuSum crosses the threshold $b$, as a function of the threshold $b$ in Fig. 9. We observe that as $b$ increases, the average number of connected components decreases, and its value is between 1.5 and 4.5. This is because for large $b$, the unaffected nodes are eliminated with high probability, and the resulting graph $G'[k]$ contains mostly the affected nodes.

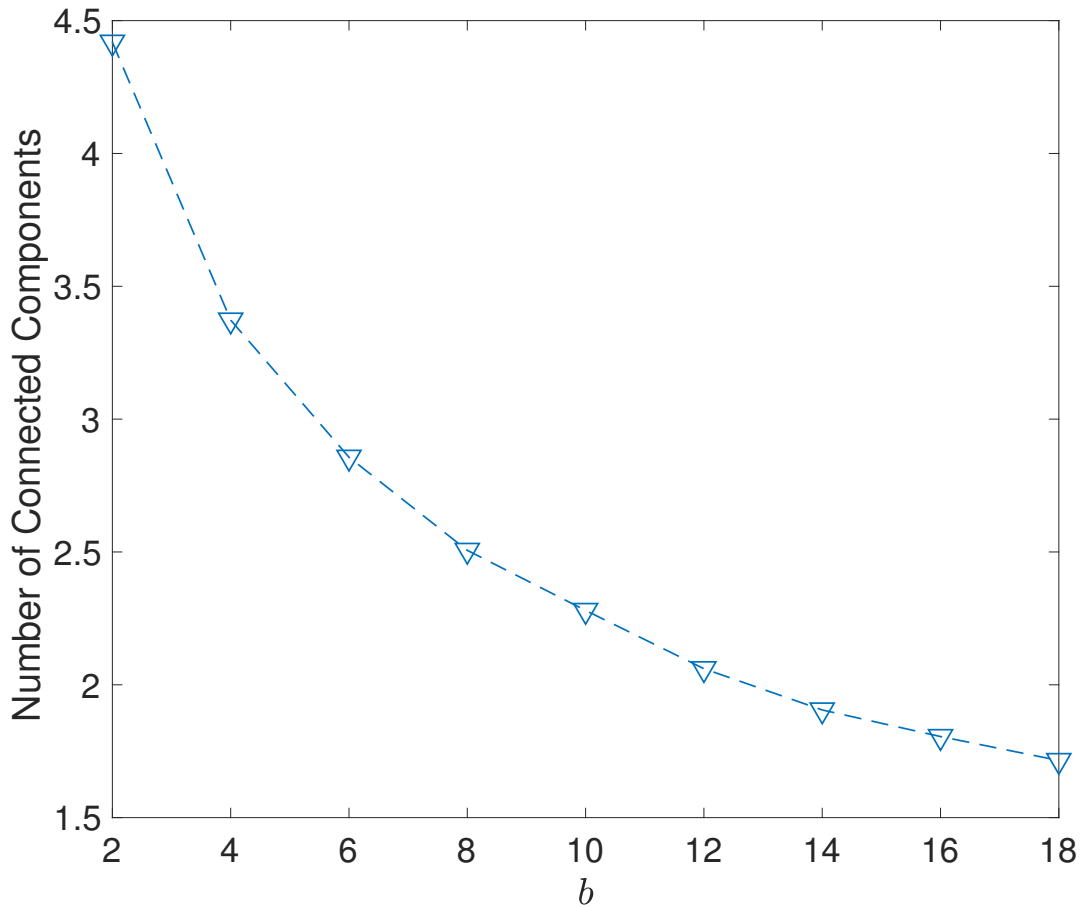

Figure 9: Number of connected components in $G'[k]$ when N-CuSum crosses the threshold.

Lastly, we compare the computational complexity of these four algorithms. We consider the lattice network in Fig. 7, and all nodes are not affected. We run these algorithms for 10000 steps without stopping ($b = \infty$), and repeat the experiment for 100 times. We run the experiment on a 2.0GHz Intel core i5 CPU using Matlab. The average time consumption for 10000 steps (in seconds) is given in Table 1. We observe that these four algorithms run very fast with 10000 steps; in particular, they all take less than one second. Also, not surprisingly, the algorithms that exploit the network structure consume more computational power.

Table 1: Comparison of time consumption

| Generalized Multi-chart CuSum | Network Generalized Multi-chart CuSum | S-CuSum | N-CuSum |
|---|---|---|---|
| 0.15 | 0.77 | 0.12 | 0.94 |

# 7 Discussion

The results in this paper can be easily generalized to the case in which the distributions of the samples are different across the different nodes (heterogeneous sensors). For example, the generalized S-CuSum algorithm for this case is also constructed by comparing the sum of the smallest $L-\eta+1$ local CuSums to a threshold $b$, and hence can be implemented efficiently. The asymptotic optimality of this algorithm can also be established similarly, but the optimal performance takes on a more complicated form.

In our modeling and analysis we have assumed that only one connected subgraph in the network is affected. In an arbitrarily connected network, if two or more events occur simultaneously, two or more connected subgraphs could be affected. Here the goal might be to detect whether there exists *one* event that has affected at least $\eta$ nodes; two small events are not of interest. Then the S-CuSum algorithm is clearly not going to be asymptotically optimal in this case, because it clubs all the affected nodes together ignoring the network structure. The N-CuSum algorithm is also not going to be asymptotically optimal for the following reason. The N-CuSum algorithm may not be able to correctly delete the unaffected nodes, and this could result in bridges between small affected connected subgraphs (with less than $\eta$ affected nodes) to form a big one (with more than $\eta$ nodes), thus triggering false alarms. Developing an efficient algorithm to handle the case of two or more events occurring simultaneously is an interesting open problem.

In this paper, the observations are assumed to be i.i.d. before and after the change-point at each node. It is clearly of interest to generalize to the case with non i.i.d. observations using tools described in [5]. Furthermore, we assumed that the data generating distributions before and after a node is affected by the event are known. In many practical applications, this assumption may not hold, and it is of further interest to construct algorithms that do not rely on complete knowledge of the distributions and still provide good performance [39,40]. In this paper, we consider a centralized setting, in which all the samples are available at a fusion center. It is also of practical interest to extend our results to the distributed setting where this is no fusion center and the sensors communicate directly with each other. The adversarial setting is also worth exploring, in which some nodes may be comprised by an adversarial party.

## Acknowledgement

The authors would like to thank Dr. Ananthram Swami and Dr. Georgios Fellouris for valuable discussions and comments.

# A  A Useful Lemma

We recall the following useful lemma, which is a slight generalization of the Weak Law of Large Numbers.

**Lemma 1.** *[41, Lemma A.1] Suppose random variables $Y_1, Y_2, \ldots, Y_k$ are i.i.d. on $(\Omega, \mathcal{F}, \mathbb{P})$ with $\mathbb{E}[Y_i] = \mu > 0$, and denote $S_k = \sum_{i=1}^{k} Y_i$, then for any $\epsilon > 0$, as $n \to \infty$,*

$$\mathbb{P}\left(\frac{\max_{1 \leq k \leq n} S_k}{n} - \mu > \epsilon\right) \to 0. \tag{42}$$

# B  Proof of Theorem 2

For a given tuple of $\{d_\eta, d_{\eta+1}, \ldots, d_{L-1}\}$, it can be shown that

$$\begin{aligned} J_{\boldsymbol{D}}[\tau] &= \sup_{\nu_1 \leq \cdots \leq \nu_\eta < \infty} \mathbb{E}_{\boldsymbol{\nu}}[\tau - \nu_\eta | \tau \geq \nu_\eta] \\ &\geq \sup_{\substack{\nu_1 = \nu_2 = \cdots = \nu_{\eta-1} = 1 \\ 1 \leq \nu_\eta < \infty}} \mathbb{E}_{\boldsymbol{\nu}}[\tau - \nu_\eta | \tau \geq \nu_\eta]. \end{aligned} \tag{43}$$

It then suffices to lower bound (43) for any stopping rule $\tau$ that satisfies the false alarm constraint. In the following of the proof, $\boldsymbol{\nu}$ is specified by $\nu_1 = \nu_2 = \cdots = \nu_{\eta-1} = 1$, $\nu_\eta$, and $d_\eta, d_{\eta+1}, \ldots, d_{L-1}$.

For simplicity, for any $\epsilon > 0$, denote

$$\alpha_\gamma = \log \gamma \left( \sum_{i=1}^{h-1} \frac{c_i}{iI} + \frac{1 - \sum_{i=1}^{h-1} c_i}{hI} \right) (1 - \epsilon). \tag{44}$$

By the Markov inequality,

$$\begin{aligned} &\mathbb{E}_{\boldsymbol{\nu}}[\tau - \nu_\eta | \tau \geq \nu_\eta] \\ &\geq \mathbb{P}_{\boldsymbol{\nu}}(\tau - \nu_\eta \geq \alpha_\gamma | \tau \geq \nu_\eta) \alpha_\gamma. \end{aligned} \tag{45}$$

It then suffices to show that

$$\mathbb{P}_{\boldsymbol{\nu}}(\tau - \nu_\eta \geq \alpha_\gamma | \tau \geq \nu_\eta) \to 1, \tag{46}$$

as $\gamma \to \infty$.

We denote $\bar{\boldsymbol{\nu}} = \{1, \ldots, 1, \infty, \ldots, \infty\}$ with the first $\eta - 1$ elements being 1, and all the remaining elements being infinity. Clearly, under $\mathbb{P}_{\bar{\boldsymbol{\nu}}}$, there are $\eta - 1$ affected nodes. For any stopping time $\tau$ that satisfies the false alarm constraint, we have

$$\mathbb{E}_{\bar{\boldsymbol{\nu}}}[\tau] \geq \gamma, \tag{47}$$

which implies that for each $m < \gamma$, there exists some $\nu \geq 1$, such that

$$\mathbb{P}_{\bar{\boldsymbol{\nu}}}(\tau \geq \nu) > 0 \text{ and } \mathbb{P}_{\bar{\boldsymbol{\nu}}}(\tau < \nu + m | \tau \geq \nu) \leq \frac{m}{\gamma}. \tag{48}$$

This can be shown by contradiction as in [22, Theorem 1].

By a change of measure argument, it follows that

$$
\begin{aligned}
&\mathbb{P}_{\bar{\boldsymbol{\nu}}}\left(\nu_\eta \leq \tau < \nu_\eta + \alpha_\gamma\right) \\
&= \mathbb{E}_{\bar{\boldsymbol{\nu}}}\left(\mathbb{1}_{\{\nu_\eta \leq \tau < \nu_\eta + \alpha_\gamma\}}\right) \\
&= \mathbb{E}_{\boldsymbol{\nu}}\left(\mathbb{1}_{\{\nu_\eta \leq \tau < \nu_\eta + \alpha_\gamma\}} \frac{\mathbb{P}_{\bar{\boldsymbol{\nu}}}(\boldsymbol{X}[\nu_\eta, \tau])}{\mathbb{P}_{\boldsymbol{\nu}}(\boldsymbol{X}[\nu_\eta, \tau])}\right) \\
&\geq \mathbb{E}_{\boldsymbol{\nu}}\left(\mathbb{1}_{\{\nu_\eta \leq \tau < \nu_\eta + \alpha_\gamma, \log \frac{\mathbb{P}_{\bar{\boldsymbol{\nu}}}(\boldsymbol{X}[\nu_\eta, \tau])}{\mathbb{P}_{\boldsymbol{\nu}}(\boldsymbol{X}[\nu_\eta, \tau])} \geq -a\}} \frac{\mathbb{P}_{\bar{\boldsymbol{\nu}}}(\boldsymbol{X}[\nu_\eta, \tau])}{\mathbb{P}_{\boldsymbol{\nu}}(\boldsymbol{X}[\nu_\eta, \tau])}\right) \\
&\geq e^{-a} \mathbb{P}_{\boldsymbol{\nu}}\left(\nu_\eta \leq \tau < \nu_\eta + \alpha_\gamma, \log \frac{\mathbb{P}_{\bar{\boldsymbol{\nu}}}(\boldsymbol{X}[\nu_\eta, \tau])}{\mathbb{P}_{\boldsymbol{\nu}}(\boldsymbol{X}[\nu_\eta, \tau])} \geq -a\right) \\
&= e^{-a} \mathbb{P}_{\boldsymbol{\nu}}\left(\nu_\eta \leq \tau < \nu_\eta + \alpha_\gamma, \log \frac{\mathbb{P}_{\boldsymbol{\nu}}(\boldsymbol{X}[\nu_\eta, \tau])}{\mathbb{P}_{\bar{\boldsymbol{\nu}}}(\boldsymbol{X}[\nu_\eta, \tau])} \leq a\right) \\
&\geq e^{-a} \mathbb{P}_{\boldsymbol{\nu}}\bigg(\nu_\eta \leq \tau < \nu_\eta + \alpha_\gamma, \\
&\qquad\qquad\qquad \max_{\nu_\eta \leq j \leq \nu_\eta + \alpha_\gamma} \log \frac{\mathbb{P}_{\boldsymbol{\nu}}(\boldsymbol{X}[\nu_\eta, j])}{\mathbb{P}_{\bar{\boldsymbol{\nu}}}(\boldsymbol{X}[\nu_\eta, j])} \leq a\bigg),
\end{aligned}
\tag{49}
$$

where $a$ will be specified later.

The event $\{\tau \geq \nu_\eta\}$ only depends on $\boldsymbol{X}[1, \nu_\eta - 1]$, which follows the same distribution under $\mathbb{P}_{\boldsymbol{\nu}}$ and $\mathbb{P}_{\bar{\boldsymbol{\nu}}}$. This implies that

$$
\mathbb{P}_{\boldsymbol{\nu}}(\tau \geq \nu_\eta) = \mathbb{P}_{\bar{\boldsymbol{\nu}}}(\tau \geq \nu_\eta).
\tag{50}
$$

It then follows that

$$
\begin{aligned}
&\mathbb{P}_{\bar{\boldsymbol{\nu}}}\left(\nu_\eta \leq \tau < \nu_\eta + \alpha_\gamma | \tau \geq \nu_\eta\right) \\
&\geq e^{-a} \mathbb{P}_{\boldsymbol{\nu}}\bigg(\nu_\eta \leq \tau < \nu_\eta + \alpha_\gamma, \\
&\qquad\qquad \max_{\nu_\eta \leq j \leq \nu_\eta + \alpha_\gamma} \log \frac{\mathbb{P}_{\boldsymbol{\nu}}(\boldsymbol{X}[\nu_\eta, j])}{\mathbb{P}_{\bar{\boldsymbol{\nu}}}(\boldsymbol{X}[\nu_\eta, j])} \leq a \bigg| \tau \geq \nu_\eta\bigg).
\end{aligned}
\tag{51}
$$

Due to the fact that for any events $A$ and $B$, $\mathbb{P}(A \cap B) \geq \mathbb{P}(A) - \mathbb{P}(B^c)$, it follows that

$$
\begin{aligned}
&\mathbb{P}_{\boldsymbol{\nu}}\bigg(\nu_\eta \leq \tau < \nu_\eta + \alpha_\gamma, \\
&\qquad\qquad \max_{\nu_\eta \leq j \leq \nu_\eta + \alpha_\gamma} \log \frac{\mathbb{P}_{\boldsymbol{\nu}}(\boldsymbol{X}[\nu_\eta, j])}{\mathbb{P}_{\bar{\boldsymbol{\nu}}}(\boldsymbol{X}[\nu_\eta, j])} \leq a \bigg| \tau \geq \nu_\eta\bigg) \\
&\geq \mathbb{P}_{\boldsymbol{\nu}}\left(\nu_\eta \leq \tau < \nu_\eta + \alpha_\gamma \bigg| \tau \geq \nu_\eta\right) \\
&\quad - \mathbb{P}_{\boldsymbol{\nu}}\bigg(\max_{\nu_\eta \leq j \leq \nu_\eta + \alpha_\gamma} \log \frac{\mathbb{P}_{\boldsymbol{\nu}}(\boldsymbol{X}[\nu_\eta, j])}{\mathbb{P}_{\bar{\boldsymbol{\nu}}}(\boldsymbol{X}[\nu_\eta, j])} > a \bigg| \tau \geq \nu_\eta\bigg) \\
&\stackrel{(a)}{=} \mathbb{P}_{\boldsymbol{\nu}}\left(\nu_\eta \leq \tau < \nu_\eta + \alpha_\gamma \bigg| \tau \geq \nu_\eta\right) \\
&\quad - \mathbb{P}_{\boldsymbol{\nu}}\bigg(\max_{\nu_\eta \leq j \leq \nu_\eta + \alpha_\gamma} \log \frac{\mathbb{P}_{\boldsymbol{\nu}}(\boldsymbol{X}[\nu_\eta, j])}{\mathbb{P}_{\bar{\boldsymbol{\nu}}}(\boldsymbol{X}[\nu_\eta, j])} > a\bigg),
\end{aligned}
\tag{52}
$$

where $(a)$ is due to the fact that the event $\{\tau \geq \nu_\eta\}$ only depends on $\boldsymbol{X}[1, \nu_\eta - 1]$, which is independent from $\boldsymbol{X}[\nu_\eta, j], \forall \nu_\eta \leq j \leq \nu_\eta + \alpha_\gamma$.

Combining (51) and (52), we obtain

$$\mathbb{P}_{\boldsymbol{\nu}}\left(\nu_\eta \leq \tau < \nu_\eta + \alpha_\gamma \Big| \tau \geq \nu_\eta\right)$$
$$\leq e^a \mathbb{P}_{\bar{\boldsymbol{\nu}}}\left(\nu_\eta \leq \tau < \nu_\eta + \alpha_\gamma | \tau \geq \nu_\eta\right)$$
$$+ \mathbb{P}_{\boldsymbol{\nu}}\left(\max_{\nu_\eta \leq j \leq \nu_\eta + \alpha_\gamma} \log \frac{\mathbb{P}_{\boldsymbol{\nu}}(\boldsymbol{X}[\nu_\eta, j])}{\mathbb{P}_{\bar{\boldsymbol{\nu}}}(\boldsymbol{X}[\nu_\eta, j])} > a\right). \tag{53}$$

By (48), it follows that for $m = \alpha_\gamma$, there exists $\nu_\eta$, such that

$$\mathbb{P}_{\bar{\boldsymbol{\nu}}}\left(\nu_\eta \leq \tau < \nu_\eta + \alpha_\gamma \Big| \tau \geq \nu_\eta\right) \leq \frac{\alpha_\gamma}{\gamma}. \tag{54}$$

Let $a = (1 - \epsilon^2) \log \gamma$, then

$$e^a \mathbb{P}_{\bar{\boldsymbol{\nu}}}\left(\nu_\eta \leq \tau < \nu_\eta + \alpha_\gamma | \tau \geq \nu_\eta\right)$$
$$\leq \gamma^{1-\epsilon^2} \frac{\alpha_\gamma}{\gamma}$$
$$\to 0, \text{ as } \gamma \to \infty. \tag{55}$$

We then show the second term in (52) also converges to 0 as $\gamma \to \infty$. It can be shown that

$$\mathbb{P}_{\boldsymbol{\nu}}\left(\max_{\nu_\eta \leq j \leq \nu_\eta + \alpha_\gamma} \log \frac{\mathbb{P}_{\boldsymbol{\nu}}(\boldsymbol{X}[\nu_\eta, j])}{\mathbb{P}_{\bar{\boldsymbol{\nu}}}(\boldsymbol{X}[\nu_\eta, j])} > a\right)$$
$$= \mathbb{P}_{\boldsymbol{\nu}}\left(\max_{\nu_\eta \leq j \leq \nu_\eta + \alpha_\gamma} \sum_{i=\eta}^{|\mathcal{C}(\boldsymbol{\nu})|} \sum_{k=\nu_i}^{j} \log \frac{f_1(X_i[k])}{f_0(X_i[k])} > a\right)$$
$$\to 0, \text{ as } \gamma \to \infty, \tag{56}$$

where the last step follows by applying Lemma 1.

Combining (55) and (56), it follows that

$$\mathbb{P}_{\boldsymbol{\nu}}\left(\nu_\eta \leq \tau < \nu_\eta + \alpha_\gamma \Big| \tau \geq \nu_\eta\right) \to 0, \tag{57}$$

as $\gamma \to \infty$. This concludes the proof.

## C  Proof for Theorem 3

By the recursive structure of $(W_i[k])^+$, and the fact that it is always non-negative, and is zero when $k = 0$, then for a given $\boldsymbol{D}$, the worst-case of the average detection delay is achieved when $\nu_1 = \cdots = \nu_\eta = 1$.

Now

$$\sum_{i=\eta}^{L} \left(W_{\mu(i)}[k]\right)^+$$
$$= \min_{C':|C'|=L-\eta+1} \sum_{i \in C'} (W_i[k])^+$$
$$\stackrel{(a)}{\geq} \min_{C' \subseteq C(\boldsymbol{\nu}):|C'|=|C(\boldsymbol{\nu})|-\eta+1} \sum_{i \in C'} (W_i[k])^+, \qquad (58)$$

where $(a)$ is due to the fact that $(W_i[k])^+$ is always non-negative, especially for $i \notin C(\boldsymbol{\nu})$.

Define the stopping rule $N(b)$:

$$N(b) = \inf\left\{k : \min_{C' \subseteq C(\boldsymbol{\nu}):|C'|=|C(\boldsymbol{\nu})|-\eta+1} \sum_{i \in C'} (W_i[k])^+ > b\right\}. \qquad (59)$$

It then follows that $\hat{\tau}(b) \leq N(b)$. Therefore, it suffices to establish an upper bound on the $\mathbb{E}_{\boldsymbol{\nu}}[N(b)]$.

For simplicity, for any $\epsilon > 0$, we denote

$$\alpha_b = b \left(\sum_{i=1}^{h-1} \frac{c_i}{iI} + \frac{1 - \sum_{i=1}^{h-1} c_i}{hI}\right)(1+\epsilon). \qquad (60)$$

Now

$$\mathbb{E}_{\boldsymbol{\nu}}[N(b)/\alpha_b] \leq \sum_{\ell=0}^{\infty} \mathbb{P}_{\boldsymbol{\nu}}\left(N(b)/\alpha_b > \ell\right)$$
$$\leq 1 + \sum_{\ell=1}^{\infty} \mathbb{P}_{\boldsymbol{\nu}}\left(N(b)/\alpha_b > \ell\right). \qquad (61)$$

We then bound $\mathbb{P}_{\boldsymbol{\nu}}\left(N(b)/\alpha_b > \ell\right)$ for every $\ell \geq 1$ as follows:

$$\begin{aligned}
&\mathbb{P}_{\boldsymbol{\nu}}\left(N(b)/\alpha_b > \ell\right) \\
&= \mathbb{P}_{\boldsymbol{\nu}}\left(\forall 1 \leq k \leq \ell\alpha_b : \min_{\substack{C' \subseteq C(\boldsymbol{\nu}): \\ |C'|=|C(\boldsymbol{\nu})|-\eta+1}} \sum_{i \in C'} (W_i[k])^+ < b\right) \\
&\leq \mathbb{P}_{\boldsymbol{\nu}}\bigg(\bigg(k = \alpha_b : \min_{\substack{C' \subseteq C(\boldsymbol{\nu}): \\ |C'|=|C(\boldsymbol{\nu})|-\eta+1}} \sum_{i \in C'} (W_i[k])^+ < b\bigg) \\
&\quad \bigcap \bigg(k = 2\alpha_b : \min_{\substack{C' \subseteq C(\boldsymbol{\nu}): \\ |C'|=|C(\boldsymbol{\nu})|-\eta+1}} \sum_{i \in C'} (W_i[k])^+ < b\bigg) \\
&\quad \bigcap \cdots \bigcap \bigg(k = \ell\alpha_b : \min_{\substack{C' \subseteq C(\boldsymbol{\nu}): \\ |C'|=|C(\boldsymbol{\nu})|-\eta+1}} \sum_{i \in C'} (W_i[k])^+ < b\bigg)\bigg) \\
&\stackrel{(a)}{\leq} \mathbb{P}_{\boldsymbol{\nu}}\bigg(\bigg(\min_{\substack{C' \subseteq C(\boldsymbol{\nu}): \\ |C'|=|C(\boldsymbol{\nu})|-\eta+1}} \sum_{i \in C'} Z_i[\max\{1,\nu_i\}, \alpha_b] < b\bigg) \\
&\quad \bigcap \bigg(\min_{\substack{C' \subseteq C(\boldsymbol{\nu}): \\ |C'|=|C(\boldsymbol{\nu})|-\eta+1}} \sum_{i \in C'} Z_i[\max\{\alpha_b+1,\nu_i\}, 2\alpha_b] < b\bigg) \\
&\quad \bigcap \cdots \bigcap \bigg(\min_{\substack{C' \subseteq C(\boldsymbol{\nu}): \\ |C'|=|C(\boldsymbol{\nu})|-\eta+1}} \sum_{i \in C'} Z_i[\max\{(\ell-1)\alpha_b+1,\nu_i\}, \ell\alpha_b] < b\bigg)\bigg) \\
&\stackrel{(b)}{=} \mathbb{P}_{\boldsymbol{\nu}}\bigg(\min_{\substack{C' \subseteq C(\boldsymbol{\nu}): \\ |C'|=|C(\boldsymbol{\nu})|-\eta+1}} \sum_{i \in C'} Z_i[\max\{1,\nu_i\}, \alpha_b] < b\bigg) \\
&\quad \times \mathbb{P}_{\boldsymbol{\nu}}\bigg(\min_{\substack{C' \subseteq C(\boldsymbol{\nu}): \\ |C'|=|C(\boldsymbol{\nu})|-\eta+1}} \sum_{i \in C'} Z_i[\max\{\alpha_b+1,\nu_i\}, 2\alpha_b] < b\bigg) \\
&\quad \times \cdots \times \mathbb{P}_{\boldsymbol{\nu}}\bigg(\min_{\substack{C' \subseteq C(\boldsymbol{\nu}): \\ |C'|=|C(\boldsymbol{\nu})|-\eta+1}} \sum_{i \in C'} Z_i[\max\{(\ell-1)\alpha_b+1,\nu_i\}, \ell\alpha_b] < b\bigg). \quad (62)
\end{aligned}$$

where $(a)$ is by the definition of $W_i[k]$; $(b)$ follows by the independency among the random variables: $\boldsymbol{X}[1, \alpha_b], \boldsymbol{X}[\alpha_b+1, 2\alpha_b], \ldots, \boldsymbol{X}[(\ell-1)\alpha_b+1, \ell\alpha_b]$.

We then bound the first term in (62). It follows that

$$\mathbb{P}_{\boldsymbol{\nu}}\bigg(\bigg(\min_{\substack{C' \subseteq C(\boldsymbol{\nu}): \\ |C'|=|C(\boldsymbol{\nu})|-\eta+1}} \sum_{i \in C'} Z_i[\max\{1,\nu_i\}, \alpha_b] < b\bigg)\bigg) \\
\leq \sum_{\substack{C' \subseteq C(\boldsymbol{\nu}): \\ |C'|=|C(\boldsymbol{\nu})|-\eta+1}} \mathbb{P}_{\boldsymbol{\nu}}\bigg(\sum_{i \in C'} Z_i[\max\{1,\nu_i\}, \alpha_b] < b\bigg). \quad (63)$$

It is clear that for large $b$,

$$\alpha_b > \sum_{i=1}^{h-1} d_{\eta+i-1}. \quad (64)$$

Moreover, $\sum_{i \in C'} Z_i[\max\{1, \nu_i\}, \alpha_b]$ is the summation of the log-likelihood ratios of the samples from $f_1$. Therefore, for any $C' \subseteq C(\boldsymbol{\nu})$ such that $|C'| = |C(\boldsymbol{\nu})| - \eta + 1$, $\sum_{i \in C'} Z_i[\max\{1, \nu_i\}, \alpha_b]$ is the sum of the log likelihood ratio between $f_1$ and $f_0$ of at least

$$d_\eta + 2d_{\eta+1} + \ldots + (h-1)d_{\eta+h-2} + h(\alpha_b - \sum_{i=1}^{h-1} d_{\eta+i-1}) \tag{65}$$

number of samples generated by $f_1$. Then by the Weak Law of Large Numbers, it follows that

$$\frac{\sum_{i \in C'} Z_i[\max\{1, \nu_i\}, \alpha_b]}{b} \to \beta, \tag{66}$$

in probability, where $\beta > 1$. Therefore, as $b \to \infty$,

$$\mathbb{P}_{\boldsymbol{\nu}} \left( \sum_{i \in C'} Z_i[\max\{1, \nu_i\}, \alpha_b] < b \right) \to 0. \tag{67}$$

Together with (63), this further implies that

$$\mathbb{P}_{\boldsymbol{\nu}} \left( \min_{\substack{C' \subseteq C(\boldsymbol{\nu}): \\ |C'| = |C(\boldsymbol{\nu})| - \eta + 1}} \sum_{i \in C'} Z_i[\max\{1, \nu_i\}, \alpha_b] < b \right)$$
$$\leq \sum_{\substack{C' \subseteq C(\boldsymbol{\nu}): \\ |C'| = |C(\boldsymbol{\nu})| - \eta + 1}} \delta'$$
$$\triangleq \delta, \tag{68}$$

where $\delta'$ and $\delta$ can be arbitrarily small for large $b$.

Following similar steps, we can also show that each term in (62) is upper bounded by $\delta$ for large $b$. Therefore,

$$\mathbb{P}_{\boldsymbol{\nu}} \left( N(b)/\alpha_b > \ell \right) \leq \delta^\ell, \tag{69}$$

and

$$\mathbb{E}_{\boldsymbol{\nu}}[N(b)/\alpha_b] \leq 1 + \sum_{\ell=1}^{\infty} \delta^\ell$$
$$= \frac{1}{1-\delta}. \tag{70}$$

This implies that

$$\mathbb{E}_{\boldsymbol{\nu}}[N(b)] \leq \frac{\alpha_b}{1-\delta}$$
$$= b \left( \sum_{i=1}^{h-1} \frac{c_i}{iI} + \frac{1 - \sum_{i=1}^{h-1} c_i}{hI} \right) \frac{1+\epsilon}{1-\delta}. \tag{71}$$

Due to the fact that $\epsilon$ is chosen arbitrarily and $\delta$ can be arbitrarily small for large $b$, as $b \to \infty$,

$$\mathbb{E}_{\boldsymbol{\nu}}[N(b)] \leq b \left( \sum_{i=1}^{h-1} \frac{c_i}{iI} + \frac{1 - \sum_{i=1}^{h-1} c_i}{hI} \right) (1 + o(1)). \tag{72}$$

This concludes the proof.

# D  Proof of Theorem 6

By the recursive structure of $(W_i[k])^+$, and the fact that it is always non-negative, and is zero when $k = 0$, then for a given $\boldsymbol{D}$, the worst-case of the average detection delay is achieved when $\nu_1 = \cdots = \nu_\eta = 1$.

We use the same notation of $\alpha_b$ as in (60). It can be shown that

$$\mathbb{E}_{\boldsymbol{\nu}}[\bar{\tau}(b)/\alpha_b] \leq \sum_{\ell=0}^{\infty} \mathbb{P}_{\boldsymbol{\nu}}\left(\bar{\tau}(b)/\alpha_b > \ell\right)$$

$$\leq 1 + \sum_{\ell=1}^{\infty} \mathbb{P}_{\boldsymbol{\nu}}\left(\bar{\tau}(b)/\alpha_b > \ell\right). \tag{73}$$

Recall that we assume that $\nu_1 \leq \nu_2 \leq \cdots \leq \nu_L$, and for large $b$, we have (64). We then denote $H = \{1, 2, \ldots, h\}$ as the set of indices of nodes that have changed their distribution by the time $\alpha_b$. For simplicity of notation, we use S-CuSum$_H[k]$ to denote the test statistic value of the S-CuSum on $H$ at time $k$:

$$\text{S-CuSum}_H[k] = \min_{\substack{C' \subseteq H: \\ |C'|=|H|-\eta+1}} \sum_{i \in C'} (W_i[k])^+. \tag{74}$$

We next bound $\mathbb{P}_{\boldsymbol{\nu}}\left(\bar{\tau}(b)/\alpha_b > \ell\right)$ for every $\ell \geq 1$ as follows:

$$\mathbb{P}_{\boldsymbol{\nu}}\left(\bar{\tau}(b)/\alpha_b > \ell\right)$$
$$= \mathbb{P}_{\boldsymbol{\nu}}\left(\forall 1 \leq k \leq \ell\alpha_b : \max_i \text{S-CuSum}_i[k] < b\right)$$
$$= \mathbb{P}_{\boldsymbol{\nu}}\left(\forall 1 \leq k \leq \ell\alpha_b : A[k] \cup B[k]\right)$$
$$\leq \mathbb{P}_{\boldsymbol{\nu}}\left(\forall k \in \{\alpha_b, 2\alpha_b, \ldots, \ell\alpha_b\} : A[k] \cup B[k]\right) \tag{75}$$

where $A[k]$ and $B[k]$ are two events defined as follows:

$$A[k] = \left\{\left\{\max_i \text{S-CuSum}_i[k] < b\right\} \cap \left\{\exists i : H \subseteq C_i[k]\right\}\right\},$$
$$B[k] = \left\{\left\{\max_i \text{S-CuSum}_i[k] < b\right\} \cap \left\{\forall i : H \nsubseteq C_i[k]\right\}\right\}. \tag{76}$$

We first analyze $A[k]$. Suppose that for some $i^*$, $H \subseteq C_{i^*}[k]$, then

$$\text{S-CuSum}_{i^*}[k] \geq \text{S-CuSum}_H[k]. \tag{77}$$

Combining with the fact that

$$\max_i \text{S-CuSum}_i[k] \geq \text{S-CuSum}_{i^*}[k], \tag{78}$$

it then follows that

$$\{\max_i \text{S-CuSum}_i[k] < b\} \subseteq \{\text{S-CuSum}_{i^*}[k] < b\}$$
$$\subseteq \{\text{S-CuSum}_H[k] < b\}. \tag{79}$$

This further implies that

$$A[k] \subseteq \left\{\{\text{S-CuSum}_H[k] < b\} \cap \{\exists i : H \subseteq C_i[k]\}\right\}$$
$$\subseteq \{\text{S-CuSum}_H[k] < b\}. \tag{80}$$

For any $1 \leq j \leq \ell$, it follows from the definition of CuSum statistic $W_i[k]$ that

$$\text{S-CuSum}_H[j\alpha_b]$$
$$= \min_{\substack{C' \subseteq H: \\ |C'| = |\bar{H}| - \eta + 1}} \sum_{i \in C'} (W_i[j\alpha_b])^+$$
$$\geq \min_{\substack{C' \subseteq H: \\ |C'| = |\bar{H}| - \eta + 1}} \sum_{i \in C'} Z_i[\max\{\nu_i, (j-1)\alpha_b + 1\}, j\alpha_b]. \tag{81}$$

We then define

$$A'[j\alpha_b]$$
$$= \left\{\min_{\substack{C' \subseteq H: \\ |C'| = |\bar{H}| - \eta + 1}} \sum_{i \in C'} Z_i[\max\{\nu_i, (j-1)\alpha_b + 1\}, j\alpha_b] < b\right\}. \tag{82}$$

It is clear that $\forall 1 \leq j \leq \ell$,

$$A[j\alpha_b] \subseteq A'[j\alpha_b], \tag{83}$$

and $A'[j\alpha_b]$ only depends on the samples from $(j-1)\alpha_b + 1$ to $j\alpha_b$.

We then analyze $B[k]$. By the assumption that the event propagates along the edges in the network, the sub-graph induced on $H$ is connected. If $\forall j \in H$, $W_j[k] \geq \log b$, then there must exist $C_i[k]$ that contains all nodes in $H$. Therefore, it follows that

$$\{\forall i : H \not\subseteq C_i[k]\} \subseteq \{\exists j \in H : W_j[k] < \log b\}. \tag{84}$$

Therefore, we have

$$B[k]$$
$$\subseteq \left\{\{\max_i \text{S-CuSum}_i[k] < b\} \cap \{\exists i \in H : W_i[k] < \log b\}\right\}$$
$$\subseteq \{\exists i \in H : W_i[k] < \log b\}. \tag{85}$$

Similarly, $\forall 1 \leq j \leq \ell$, it follows from the definition of $W_i[j\alpha_b]$ that

$$W_i[k] \geq Z_i[\max\{\nu_i, (j-1)\alpha_b + 1\}, j\alpha_b]. \tag{86}$$

We then define
$$B'[j\alpha_b] = \{\exists i \in H : Z_i[\max\{\nu_i, (j-1)\alpha_b + 1\}, j\alpha_b] < \log b\}. \tag{87}$$

It follows that $\forall 1 \leq j \leq \ell$,
$$B[j\alpha_b] \subseteq B'[j\alpha_b], \tag{88}$$

and $B'[j\alpha_b]$ only depends on the samples from $(j-1)\alpha_b + 1$ to $j\alpha_b$.

Combining (83) and (88), equation (75) can be further bounded as follows:
$$\begin{aligned}
&\mathbb{P}_{\boldsymbol{\nu}}\left(\bar{\tau}(b)/\alpha_b > \ell\right) \\
&\leq \mathbb{P}_{\boldsymbol{\nu}}\left(\forall k \in \{\alpha_b, 2\alpha_b, \ldots, \ell\alpha_b\} : A'[k] \cup B'[k]\right) \\
&= \prod_{j=1}^{\ell} \mathbb{P}_{\boldsymbol{\nu}}\left(A'[j\alpha_b] \cup B'[j\alpha_b]\right) \\
&\leq \prod_{j=1}^{\ell} \left(\mathbb{P}_{\boldsymbol{\nu}}\left(A'[j\alpha_b]\right) + \mathbb{P}_{\boldsymbol{\nu}}\left(B'[j\alpha_b]\right)\right).
\end{aligned} \tag{89}$$

Following similar steps as from (62) to (69), we can show that for large $b$, $\forall 1 \leq j \leq \ell$,
$$\mathbb{P}_{\boldsymbol{\nu}}\left(A'[j\alpha_b]\right) \leq \delta/2, \tag{90}$$

where $\delta$ can be arbitrarily small.

Then by the Weak Law of Large Numbers, for large $b$, we obtain
$$\begin{aligned}
&\mathbb{P}_{\boldsymbol{\nu}}\left(B'[j\alpha_b]\right) \\
&\leq \sum_{i \in H} \mathbb{P}_{\boldsymbol{\nu}}\left(Z_i[\max\{\nu_i, (j-1)\alpha_b + 1\}, j\alpha_b] < \log b\right) \\
&\leq \sum_{i \in H} \delta' \\
&\triangleq \delta/2,
\end{aligned} \tag{91}$$

where $\delta'$ can be made arbitrarily small for large $b$, and hence so can $\delta$.

Therefore,
$$\mathbb{P}_{\boldsymbol{\nu}}\left(\bar{\tau}(b)/\alpha_b > \ell\right) \leq \delta^{\ell}, \tag{92}$$

and following steps similar to those in (70) to (72), we conclude the proof.